\documentclass[preprint,12pt]{elsarticle}

\usepackage{amssymb}
\usepackage{amsmath}

\usepackage{nomencl}

\DeclareMathOperator*{\argmax}{arg\,max}

\usepackage[ruled,vlined]{algorithm2e}

\usepackage{soul}
\usepackage{comment}

\usepackage{caption}
\usepackage{soul,xcolor}
\usepackage{enumitem,kantlipsum}
\usepackage{multirow}
\usepackage{color, colortbl}
\usepackage{hyperref}
\usepackage{subcaption}
\usepackage{float}
\usepackage{subfloat}
\usepackage{ifthen}

\usepackage{nomencl}
\usepackage{xpatch}

\newlength{\nomitemorigsep}
\makenomenclature
\renewcommand{\nomgroup}[1]{%
  \itemsep\nomitemorigsep%
  \ifthenelse{%
    \equal{#1}{A}%
  }{%
  \item[\textbf{Greek Symbols}]%

  }{%
    \ifthenelse{\equal{#1}{B}}{%
    \item[\textbf{Other Variables}]%

    }{}%
  }%
  \itemsep\nomitemsep
} 

\begin{document}

\begin{frontmatter}



\title{Event-Triggered Islanding in Inverter-Based Grids \vspace{-2mm}}


\author[label1]{Ioannis Zografopoulos} 

\affiliation[label1]{organization={Engineering Department, Univesity of Massachusetts Boston},
            city={Boston},
            postcode={02125}, 
            state={MA},
            country={USA}}

\author[label2]{Charalambos Konstantinou} 

\affiliation[label2]{organization={CEMSE Division, King Abdullah University of Science and Technology},
            city={Thuwal},
            country={KSA} }

\begin{abstract}
The decentralization of modern power systems challenges the hierarchical structure of the electric grid and necessitates automated schemes to manage adverse conditions. This work proposes an adaptive isolation methodology that can divide a grid into autonomous islands, ensuring stable and economical operation amid deliberate or unintentional abnormal events. The adaptive isolation logic is event-triggered to prevent false positives, enhance detection accuracy, and reduce computational overhead. A measurement-based stable kernel representation (SKR) triggering mechanism initially inspects distributed generation controllers for abnormal behavior. The SKR then alerts an ensemble classifier to assess whether the system behavior remains within acceptable operational limits. The event-triggered adaptive isolation framework is evaluated using IEEE RTS-24 and 118-bus systems. Simulation results demonstrate that the proposed framework detects anomalous behavior with 100\% accuracy in real-time, i.e., within $22~msec$. Supply-adequate partitions are identified outperforming traditional islanding detection and formation techniques while minimizing operating costs.
\end{abstract}

    
    
    
    


\begin{keyword}
Attacks \sep detection \sep distributed generation \sep event-triggered \sep islanding \sep isolation \sep resilience.



\end{keyword}

\end{frontmatter}
\vspace{-6mm}



\vspace{-3mm}

\nomenclature[C]{$N$, $B$, $L$, $G$, $D$}{ Set of nodes, buses, lines, generators, and loads  in system.}
\nomenclature[C]{$B^{h}, B^{u}$}{ Buses in the healthy, unhealthy partitions.}
\nomenclature[C]{$L^{+}, L^{-}$}{ Set of inbound, outbound lines for a bus.}
\nomenclature[C]{$L^{\bullet}$}{ Set of uncertain lines in system.}
\nomenclature[C]{$P^{min}_{l_{i,j}}/P^{max}_{l_{i,j}}$}{ Minimum/Maximum load flow for line $l_{i,j}$.}
\nomenclature[C]{$P^{min}_{g}/P^{max}_{g}$}{ Minimum/Maximum power provided by generator $g$.}
\nomenclature[C]{$\omega_{l_{i,j}}$}{ Power flow weight of line  $l_{i,j}$.}
\nomenclature[C]{$\omega_{g}$}{ Generator capacity weight for generator  $g$.}
\nomenclature[C]{$\chi_{g}$}{ Power output growth rate for generator  $g$.}
\nomenclature[C]{$\theta_{d}$}{ Portion ($\%$) of critical loads comprising load $d$.}
\nomenclature[C]{$P^{a}_{d}$}{ Aggregated power demand by load $d$.}
\nomenclature[C]{$\lambda$}{ Weighting factor for the islanding objective function.}

\nomenclature[I]{$i, j$}{ Index of node.}
\nomenclature[I]{$l_{i,j}$}{ Line connecting nodes $i$ and $j$.}
\nomenclature[I]{$g,d$}{ Index of generator and index of load.}

\nomenclature[V]{$h_{b}, \phi_{g}, w_{l_{i,j}} $}{Binary decision variable for bus $b$, generator status, and line $l_{i,j}$.}

\nomenclature[V]{$P^{s}_{d}$}{ Power supplied to to load $d$.}
\nomenclature[V]{$\beta_{d}$}{ Portion ($\%$) of the served load demand of $d$.}
\printnomenclature

\vspace{-7mm}
\section{Introduction}
\vspace{-3mm}
In recent years, the high penetration of renewable and distributed generation (DG) has altered the structure of traditional energy systems. The considerable advantages that active and/or autonomous, e.g., microgrids (MG), distribution systems offer -- with respect to reliability, economic operation, energy loss minimization, and resilience to adverse events -- justify the thrust towards a decentralized grid architecture. The interoperability and interdependence of such architectures require careful consideration for the orchestration of DG-integrated energy
systems, especially during unexpected adverse events, e.g., faults, cyberattacks, etc. \cite{10870120, ospina2020trustworthy}.\looseness=-1

Intentional controlled islanding (ICI) has been widely used to sectionalize grid topology into healthy, supply-adequate islands during contingencies \cite{quiros2014constrained, demetriou2018real}. Various techniques have been proposed for detecting islanding conditions \cite{khamis2015islanding, najy2011bayesian, el2007data, gaing2004wavelet, lidula2012pattern, faqhruldin2014universal, khamis2016faster}, relying on remote or local measurements. Remote schemes require continuous, reliable information aggregation, making them impractical and costly due to the need for expanded communications infrastructure \cite{liu2015principal}. In contrast, local measurement-based techniques are faster, more practical, and cost-effective for detecting islanding events \cite{khamis2016faster, zografopoulos2021detection} and classifying real-time system behavior as anomalous or expected \cite{zografopoulos2021detection}.

Islanding methods based on local measurements are classified into passive, active, and hybrid approaches. Passive methods analyze system parameters such as voltage, current, frequency, and harmonic distortion \cite{li2014review}, while active methods monitor system responses to artificially injected perturbations \cite{sun2015islanding, gupta2016active}. Hybrid methods combine both approaches to improve accuracy and reduce decision uncertainty \cite{seyedi2021hybrid}. However, passive methods can be less accurate when the power mismatch is small \cite{seyedi2021hybrid}, and active methods increase costs and may degrade power quality \cite{li2014review}. Hybrid methods improve detection confidence, but they tend to have slower response times \cite{khamis2015islanding}. 

This performance trade-off has led to research that combines the advantages of passive islanding methods, such as fast detection, minimal hardware, and no grid impact, with the availability of local measurements from DG and inverter-based resources. Machine learning (ML) techniques have been used to mitigate potential drawbacks, like non-detection zones, for more accurate islanding detection. For instance, \cite{khamis2016faster} uses extreme learning machines (ELM) for event classification, while \cite{gaing2004wavelet} applies discrete wavelet transformations (DWT) and probabilistic neural networks (PNN) for decision-making. \cite{khamis2015islanding} employs phase-space transformation with PNN for fast detection, and \cite{el2007data} uses data mining and a classification and regression tree (CART) algorithm for incident classification. Pattern recognition with decision tree (DT) classifiers trained on DWT features is also used for quick detection \cite{lidula2012pattern, faqhruldin2014universal}. Additionally, \cite{najy2011bayesian} utilizes a naive Bayes classifier on data processed through the rotational invariance technique (ESPRIT) to detect islanding.
The detection techniques summarized in Table \ref{tab:IslandDecisionSpeed} leverage ML and passive islanding classifiers for fast and accurate decisions, conforming to standards like UL 1741 and IEEE 1547-2020 \cite{IEEE1547}. All methods detect islanding within the 2-second limit specified by IEEE 1547-2020. Besides speed and accuracy, computational requirements and adaptability to grid topology changes are important factors. Real-time detection on DG inverter controllers is challenging due to limited computational resources, and dynamic grid behaviors must be considered. Therefore, event-triggered, topology-agnostic algorithms using local measurements and adaptable to these changes are crucial.

\begin{table}[t]

\setlength{\tabcolsep}{1.2pt}
\centering
\vspace{-3mm}
\caption{Islanding Decision Speed and Accuracy of Machine Learning Algorithms} \vspace{-3mm}
\label{tab:IslandDecisionSpeed}
\resizebox{0.65\linewidth}{!}{
\begin{tabular}{||l|c|c|c||}
\hline \hline
\textbf{Algorithm} & \textbf{Speed} & \textbf{Accuracy} & \textbf{Reference} \\ \hline 

{Phase space transform + PNN} & {0.230 sec} & {100\%} & {\cite{khamis2015islanding}} \\  \hline 
{ESPRIT} & {0.150 sec} & {95.6\% } & {\cite{najy2011bayesian}} \\  \hline
{Classification + Regression tree} & {0.125 sec} & {94.45\%} & {\cite{el2007data}} \\  \hline
{DWT + PNN} & {0.100 sec} & {90\%} & {\cite{gaing2004wavelet}}\\  \hline 
{Decision Trees + DWT} & {0.024 sec} & {98\%} & {\cite{lidula2012pattern, faqhruldin2014universal}} \\  \hline
{Extreme learning machines} & {0.021 sec*} & {99.3\%*} & {\cite{khamis2016faster}} \\   \hline 
{\textbf{SKR trigger + SVM ensemble}} & {\textbf{0.022 sec$\dagger$} } & {\textbf{100\%} } & {\textbf{This work}} \\  \hline \hline

\end{tabular}}
\small{{\\ *Average based on reported results. $\dagger$Worst-case scenario results.}}
\vspace{-6mm}
\end{table}

\begin{figure}[t]
    \centering
     \vspace{-2mm}
    \includegraphics[width=0.8\linewidth]{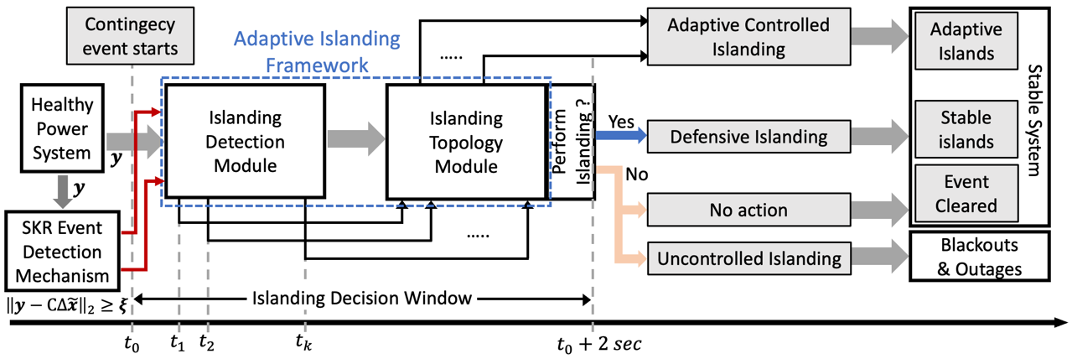}
        \vspace{-2mm}
    \caption{Overview of the event-triggered islanding framework.}
    \label{fig:framework}
    \vspace{-5mm}
\end{figure}

This work proposes an event-triggered adaptive islanding framework that comprehensively addresses the disadvantages of previous works. In Fig. \ref{fig:framework}, we present an overview of the proposed detection and isolation framework. Specifically, the first step includes the event-triggered detection of islanding conditions leveraging local measurements (i.e., passive islanding), support vector machine (SVM), and ensemble classifiers. As a result, computational-intensive ML tasks are only performed after an event has been triggered, reducing performance overheads. The second step of the framework identifies stable island partitions adaptively while minimizing costs (e.g., load shedding, uneconomic generator operation, etc.). Namely, the adaptive islanding framework is composed of two distinct computational blocks, i.e., \emph{i)} the islanding detection module and \emph{ii)} the optimal islanding topology identification module, which work in parallel (Fig. \ref{fig:framework}). 
These computational blocks work independently, and the detection module notifies the optimal islanding algorithm when an anomaly is detected. Thus, the objective of the second step of the framework is first to detect anomalous incidents and then determine supply-adequate islands while maximizing the size of the unhealthy partition. 
After the islanding, if the contingency persists, the decided islands will be adaptively segregated into smaller autonomous partitions containing the impact of the adverse event. The contributions are as follows: \vspace{-2mm}
\begin{enumerate}[wide, labelwidth=!, labelindent=0pt]
    \item An event-triggered anomaly indication scheme that identifies deliberate (e.g., cyberattacks) or accidental (e.g., faults) events. \textcolor{black}{Event detection employs a stable kernel representation (SKR) of the system using local measurements aggregated at the grid-forming DG inverters' side.} \vspace{-3mm}
    \item An ensemble classifier (after an event is triggered) to identify whether islanding should be performed. The voting classifier combines results from multiple 
    rounds to ensure correct decisions are made. \vspace{-3mm}
    \item An automated scheme provides adaptive islanding decisions that minimize operating costs and maximize the size of stable grid partitions. Case studies on the IEEE RTS-24 and 118-bus systems demonstrate the scheme's efficacy and scalability in maintaining economic and stable grid operations.
\end{enumerate}
\vspace{-2mm}
The paper is organized as follows. Section \ref{s:eventDetection} presents the event-triggered anomaly indication mechanism that prompts the islanding detection and isolation framework described in Section \ref{s:islandindDetection}. Simulation results and comparisons with other works are furnished in Section \ref{s:results} and Section \ref{s:conclusion} concludes the paper. \looseness=-1 


\vspace{-3mm}
\section{Distributed Event-triggered Anomaly Indication Mechanism} \label{s:eventDetection}
\vspace{-3mm}
In this section, we provide the preliminary definitions for the SKR extraction. 
We also elucidate the operation of the SKR event detection mechanism which supports the event-triggered functionality of our adaptive isolation. 

\vspace{-3mm}
\subsection{Distributed Generation System Model Dynamics}
According to the mathematical modeling in \cite{liu2014impact}, the small signal model of an inverter-based DG subsystem can be represented as: 
\vspace{-2mm}
\begin{equation}\vspace{-2mm}
\mathbf{E} \Delta \dot{\mathbf{x}}=\mathbf{A} \Delta \mathbf{x}+\mathbf{Fr}_{0}
\label{eq:statespace}
\end{equation}

\noindent where $\mathbf{A}$, $\mathbf{E}$, and $\mathbf{F}$ are the system and parameter (singular) matrices, $\mathbf{x}=[\delta, \omega,\mathbf{i}_{\mathbf{d}}, \mathbf{i}_{\mathbf{q}}, \mathbf{i}_{\mathrm{dref}}, \mathbf{i}_{\mathrm{qref}}, \mathbf{v}_{\mathbf{d}}, \mathbf{v}_{\mathbf{q}}, \mathbf{P}, \mathbf{Q},$
\noindent$ \mathbf{V}_{\mathbf{d}}, \mathbf{V}_{\mathbf{q}}, \mathbf{i}_{\mathbf{x}}, \mathbf{i}_{\mathbf{y}},\mathbf{V}_{\mathbf{x}}, \mathbf{V}_{\mathbf{y}}]^{T}$, and $\mathbf{r}_{0}=\left[\omega_{0}\right]^{T}$, respectively. 
$\mathbf{E}$ and $\mathbf{A}$ are sparse matrices of the following forms and are defined in Appendix A of \cite{liu2014impact}. \vspace{-2mm}
\vspace{-1mm}
\begin{equation}
\mathbf{E}=\left[\begin{array}{cc}
\mathbf{E}_{1} & \mathbf{E}_{2} \\
0 & 0
\end{array}\right]
\textsf{~~and~~} 
\mathbf{A}=\left[\begin{array}{ll}
\mathbf{A}_{1} & \mathbf{A}_{2} \\
\mathbf{A}_{3} & \mathbf{A}_{4}
\end{array}\right]
\end{equation}

According to \cite{zografopoulos2021detection} the state variable vector can be described as $\tilde{\mathbf{x}}=\left[\Delta \delta, \Delta \omega, \Delta \mathbf{i}_{\mathbf{d}}, \Delta \mathbf{i}_{\mathbf{q}}, \Delta \mathbf{i}_{\mathrm{dref}}, \Delta \mathbf{i}_{\mathbf{q r e f}}, \Delta \mathbf{u}_{\mathbf{d}}, \Delta \mathbf{u}_{\mathbf{q}}\right]^{T}$, which leads to a state space system equivalent of the following form:
\vspace{-2mm}
\begin{equation}
\Delta \dot{\tilde{\mathbf{x}}}=\tilde{\mathbf{A}} \Delta \tilde{\mathbf{x}}
\label{eq:newstatespace}
\end{equation}
\noindent where $\tilde{\mathbf{x}} = [\tilde{\mathbf{x}}_{1}, \dots, \tilde{\mathbf{x}}_{m}]^{T}$ and 
\vspace{-1mm}
\begin{equation}
\tilde{\mathbf{A}}=\left(\mathbf{E}_{1}-\mathbf{E}_{2} \mathbf{A}_{4}^{-1} \mathbf{A}_{3}\right)^{-1}\left(\mathbf{A}_{1}-\mathbf{A}_{2} \mathbf{A}_{4}^{-1} \mathbf{A}_{3}\right)
\end{equation}

\noindent The feedback controller of the system defined in \eqref{eq:newstatespace} can be defined as: \vspace{-2mm}
\begin{equation}
\left\{\begin{array}{l}
\mathbf{u}=\mathbf{K y} \\
\mathbf{y}=\mathbf{C} \Delta \tilde{\mathbf{x}}
\end{array}\right.
\label{eq:secondarymixspace}
\end{equation}

\noindent with $\mathbf{u} = [u_{1}, \dots, u_{n}]^{T}$ the control inputs, $\mathbf{y} = [y_{1}, \dots, y_{n}]^{T}$ the output measurement vectors, and $\mathbf{C} \in\mathbb{R}^{n\times m}$ being the output matrix. Ignoring system and process measurement noise, and combining Eqs. \eqref{eq:statespace} - \eqref{eq:secondarymixspace} the system model can be described as:
\vspace{-2mm}
\begin{equation}
\left\{\begin{array}{l}
\Delta \dot{\tilde{\mathbf{x}}}=\tilde{\mathbf{A}} \Delta \tilde{\mathbf{x}}+\mathbf{B u} \\
\mathbf{y}=\mathbf{C} \Delta \tilde{\mathbf{x}}
\end{array}\right.
\label{eq:statespacecontrolobserver}
\end{equation}

The calculation of state space residuals (from Eq. \eqref{eq:statespacecontrolobserver}) can then be used for the detection of attacks, and data integrity violations \cite{zografopoulos2021detection, 8894512}. 

\begin{figure}[t]
    \centering
    \includegraphics[width=0.3\linewidth]{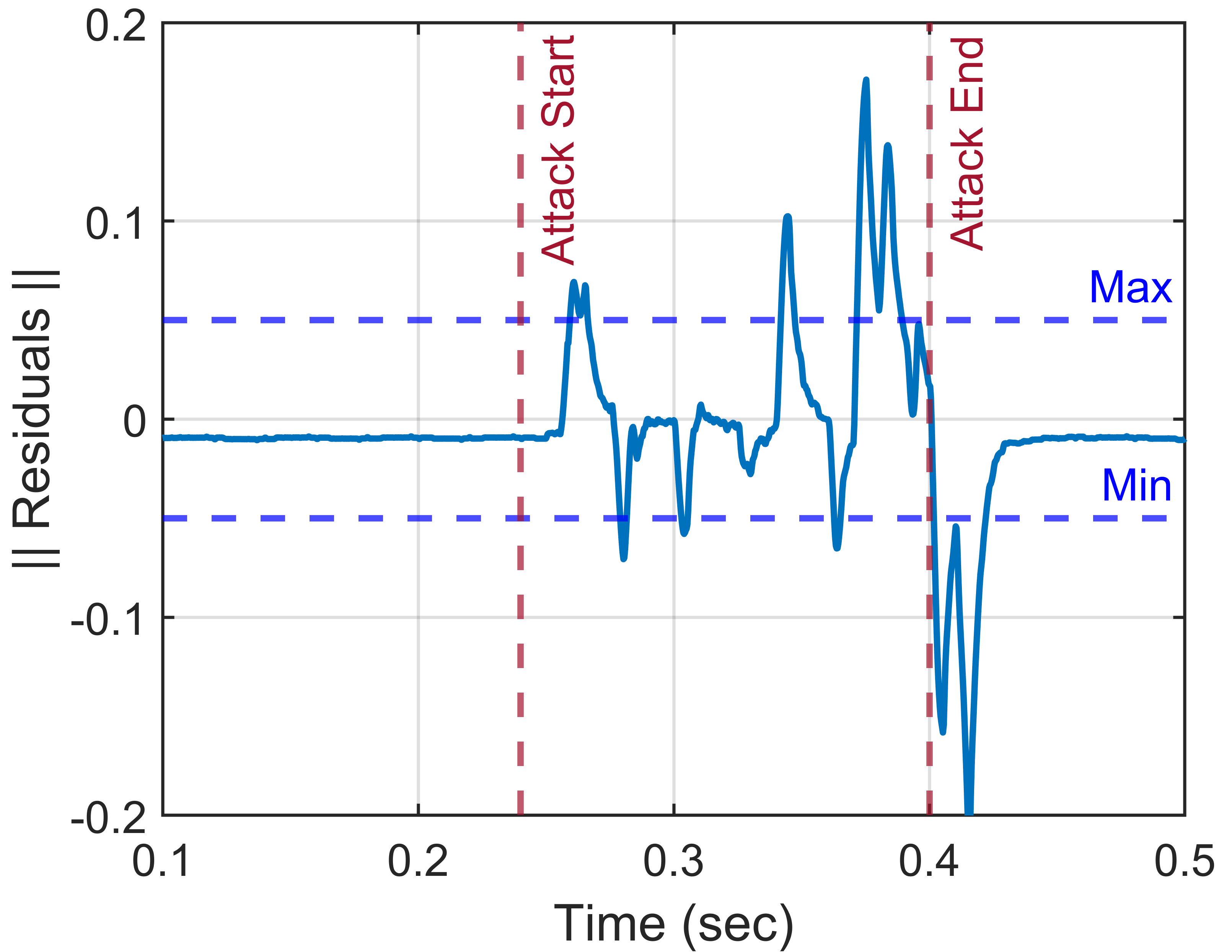}
    \caption{Response of the SKR event detection scheme during a control input attack targeting DG inverter controller.}
    \label{fig:SKR_alarm}
    \vspace{-5mm}
\end{figure}

\begin{figure*}[t]
\centering
    \subfloat[]{
        \includegraphics[width=0.24\textwidth]{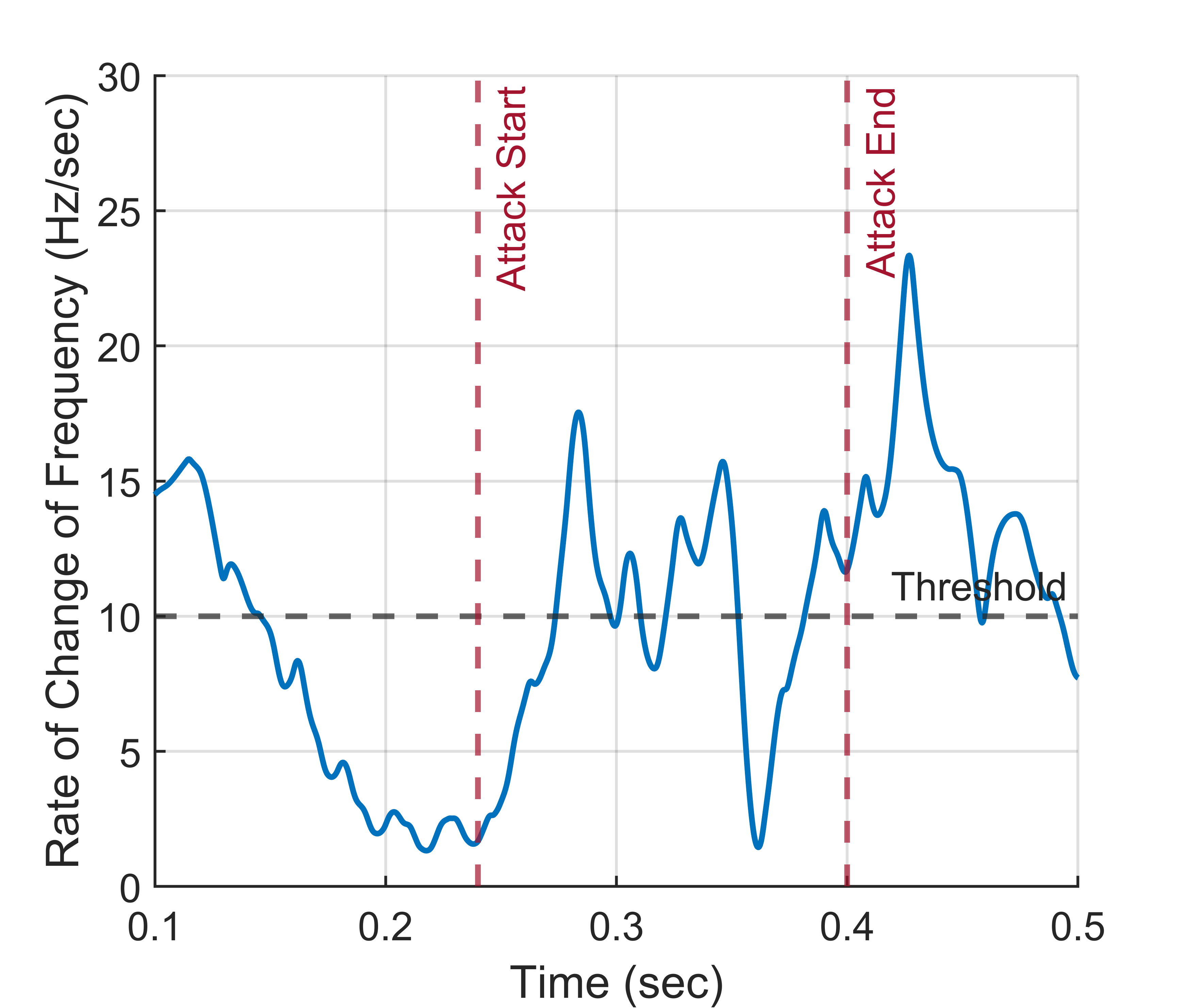}
        \label{fig:ROCOF}
    } 
    \subfloat[]{
        \includegraphics[width=0.24\linewidth]{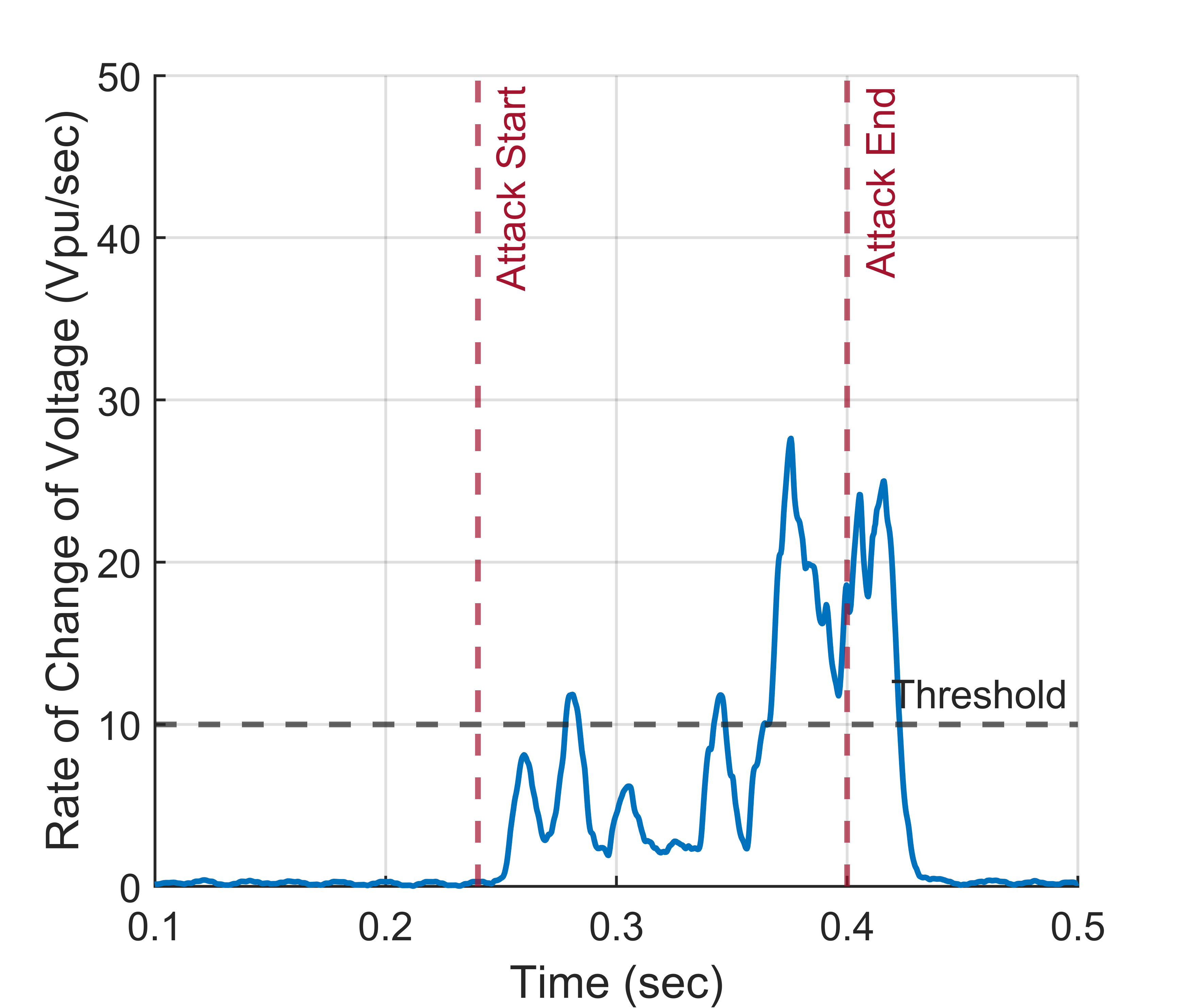}
        \label{fig:ROCOV}
    } 
    \subfloat[]{
        \includegraphics[width=0.24\linewidth]{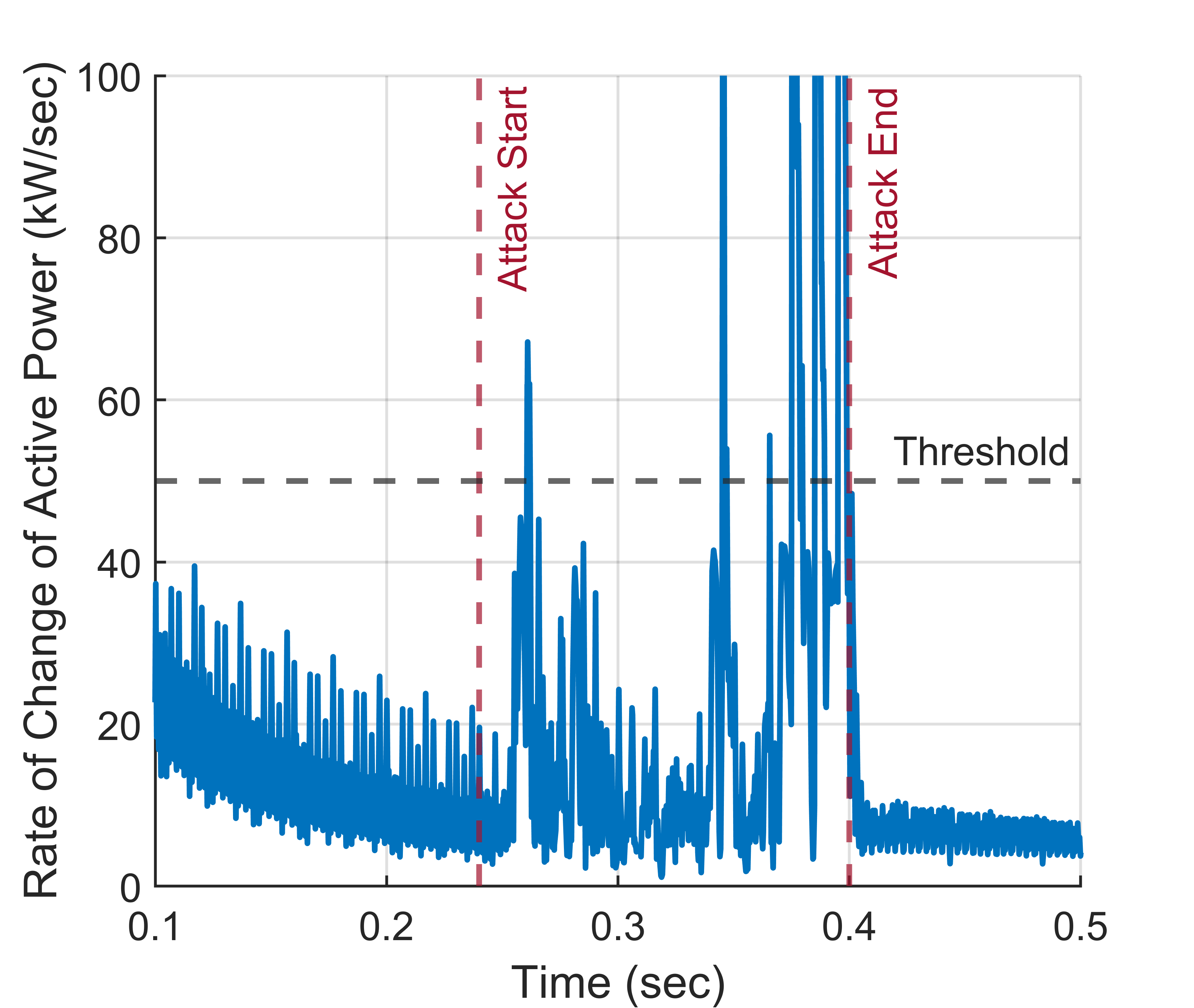}
        \label{fig:ROCOAP}
    } 
    \subfloat[]{
        \includegraphics[width=0.24\linewidth]{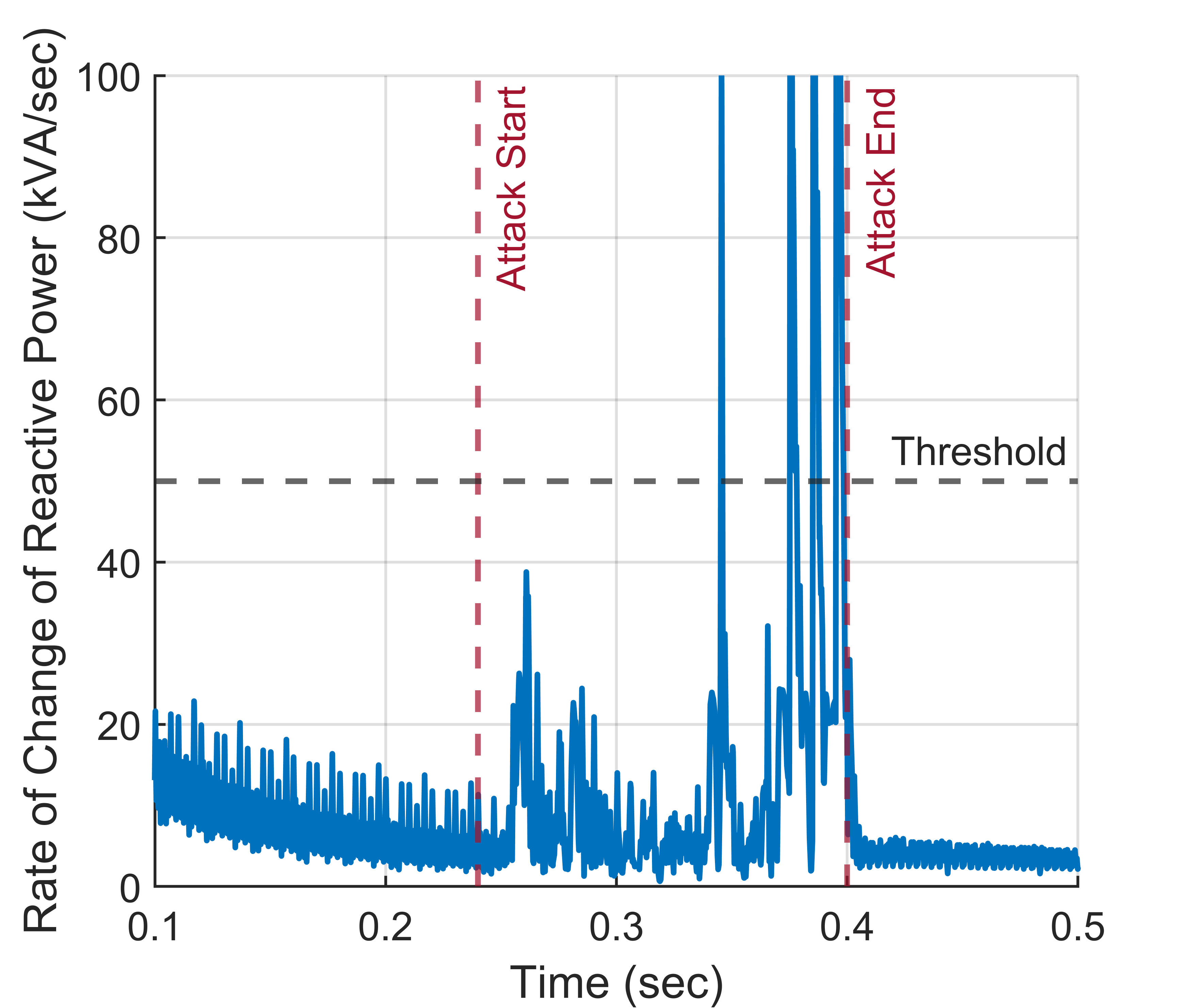}
        \label{fig:ROCORP}
    } \\
\vspace{-3mm}    
\caption[CR]{Response of passive islanding detection schemes during a control input attack targeting DG 2, rate of change of: \subref{fig:ROCOF}) frequency (ROCOF), \subref{fig:ROCOV}) voltage (ROCOV), \subref{fig:ROCOAP}) active power (ROCOAP), and \subref{fig:ROCORP}) reactive power (ROCORP).} 
\vspace{-4mm}
\label{fig:detectors}
\end{figure*}

\begin{figure}[t]
    \centering
    \includegraphics[width=0.9\linewidth]{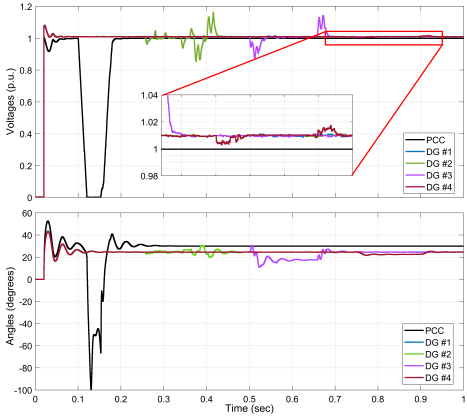}
    \vspace{-2mm}
    \caption{Voltage and angles of the system during different adverse scenarios.}
    \label{fig:voltAngles}
    \vspace{-5mm}
\end{figure}

\vspace{-4mm}
\subsection{Event-triggered Stable Kernel Representation Alarms} \label{s:SKRalarms}

 State residuals, $\mathbf{r}$, are defined as the deviation of real measurements $\mathbf{y}$ and their corresponding estimated values $\Delta \tilde{\mathbf{x}}$ for the same system. Then leveraging a $\chi^2$-distribution with predefined degrees of freedom (e.g., system states) and confidence intervals (e.g., values within the 95\% or 99\% confidence interval) as indicated by the system model, we can define a static threshold $\xi$. If the residual deviation is greater than $\xi$, i.e., $\mathbf{r}=\| \mathbf{y}-\mathbf{C} \Delta \tilde{\mathbf{x}} \|_{2} > \xi$, then the measurements are considered corrupted. \textcolor{black}{Building on our previous work  \cite{zografopoulos2021detection}, our scheme leverages state residual deviations as an \emph{anomaly indicator}. Instead of using static predefined thresholds $\xi_{i}$ for each subsystem, we leverage system measurements to develop constantly updating dynamic thresholds that serve as inputs to the islanding detection module. 
Once a violation of the state residuals distribution is detected, then control is handed over to the 
islanding detection module (Section \ref{s:islandindDetection}), which is responsible for identifying if a compromise or any other adverse event has occurred.} This two-step approach reduces potential false positive alarms and conserves computation bandwidth since the distributed agents do not constantly monitor for anomalous events. \looseness=-1 

\textcolor{black}{We illustrate the effectiveness of SKR in generating reliable alarm signals during anomalous conditions and its improvement over traditional passive detection methods in Fig. \ref{fig:SKR_alarm} and \ref{fig:detectors}. The event detection is performed in the Canadian urban distribution feeder system benchmark, which comprises four DGs that, in our case, are inverter-based \cite{liu2014impact, zografopoulos2022time}.} We employ a control input attack affecting the operation of a DG inverter controller as an illustrative example. The control input attack starts at $0.25$ $sec$ and finishes at $0.4$ $sec$. In Fig. \ref{fig:SKR_alarm}, the SKR triggering mechanism  generates multiple abnormal event signals throughout the duration of the control input attack, proving its efficacy. On the other hand, as demonstrated in Fig. \ref{fig:detectors}, even though some passive islanding techniques, e.g., rate-of-change-of-frequency (ROCOF), rate-of-change-of-voltage (ROCOV), rate-of-change-of-active/reactive power (ROCOAP/ROCORP) could also detect  anomalous behavior, their detection speed and robustness (values exceeding thresholds) are significantly worse than the SKR that produced multiple alarms within this time frame. Furthermore, metrics like ROCOF could still keep providing false alarms even after the attack has stopped, i.e., $t\geq 0.4$ $sec$. 

\begin{figure*}[t]
\centering
    \subfloat[]{
        \includegraphics[width=0.23\textwidth]{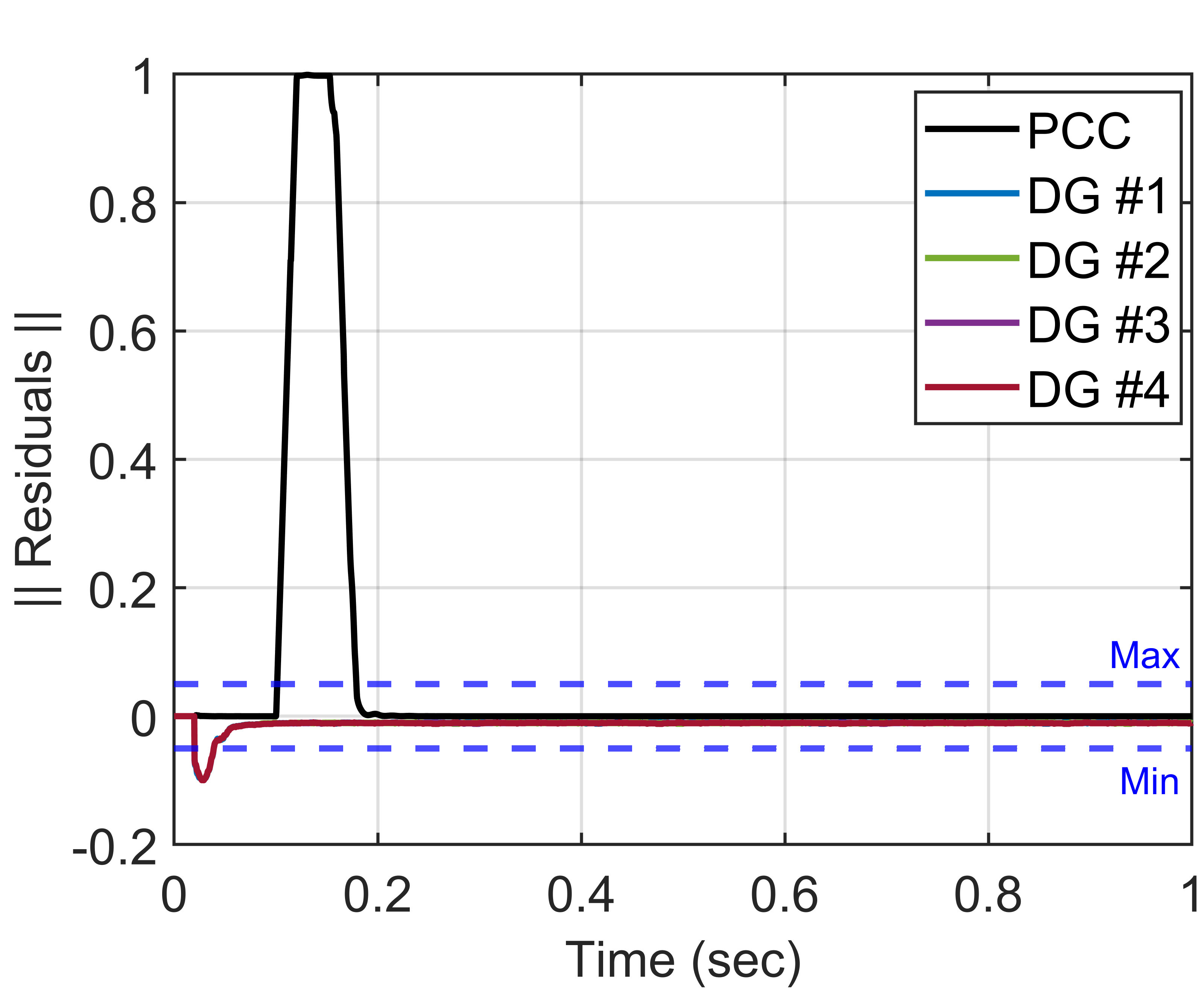}
        \label{fig:PCC}
    } 
    \subfloat[]{
        \includegraphics[width=0.23\linewidth]{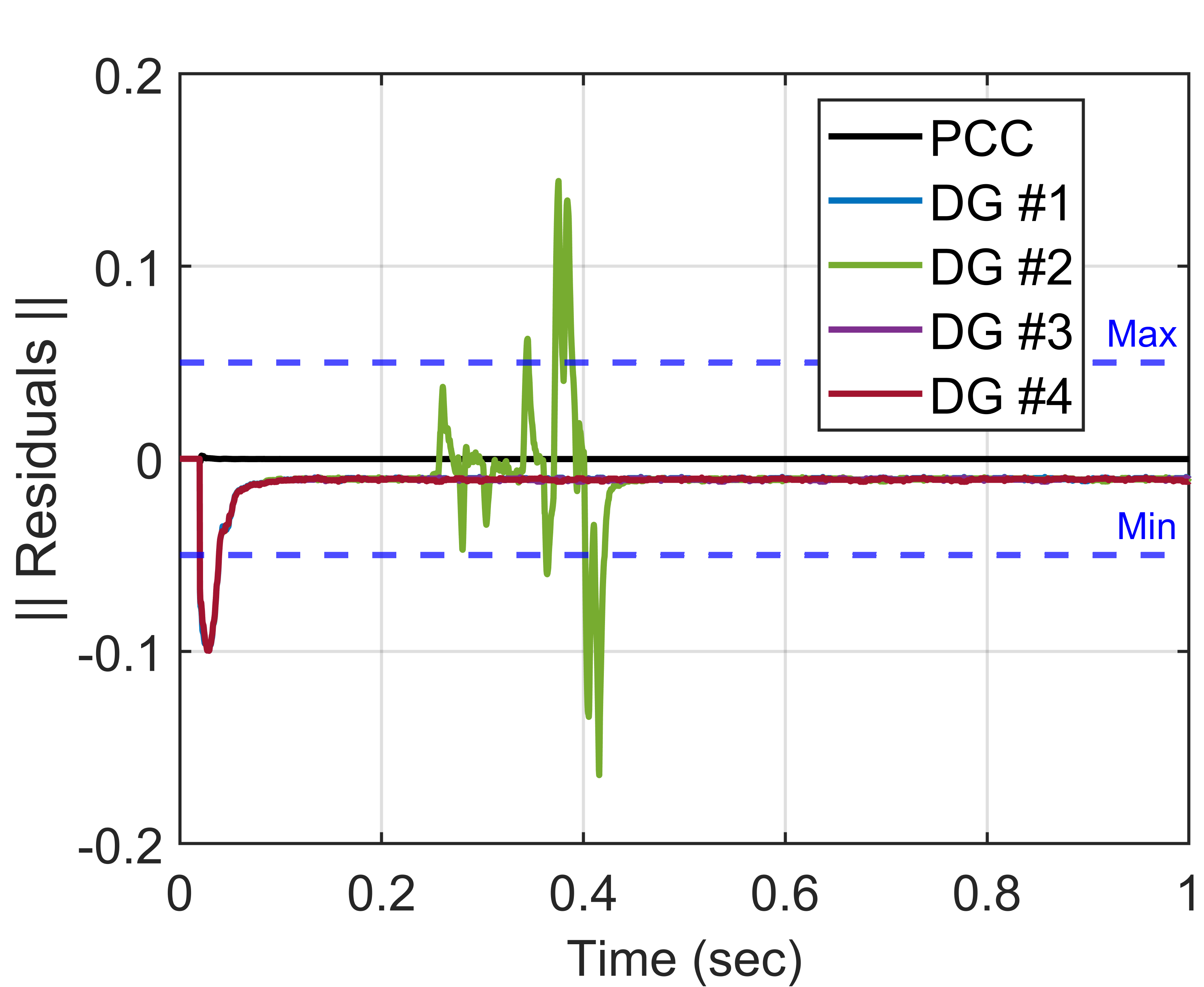}
        \label{fig:control}
    } 
    \subfloat[]{
        \includegraphics[width=0.23\linewidth]{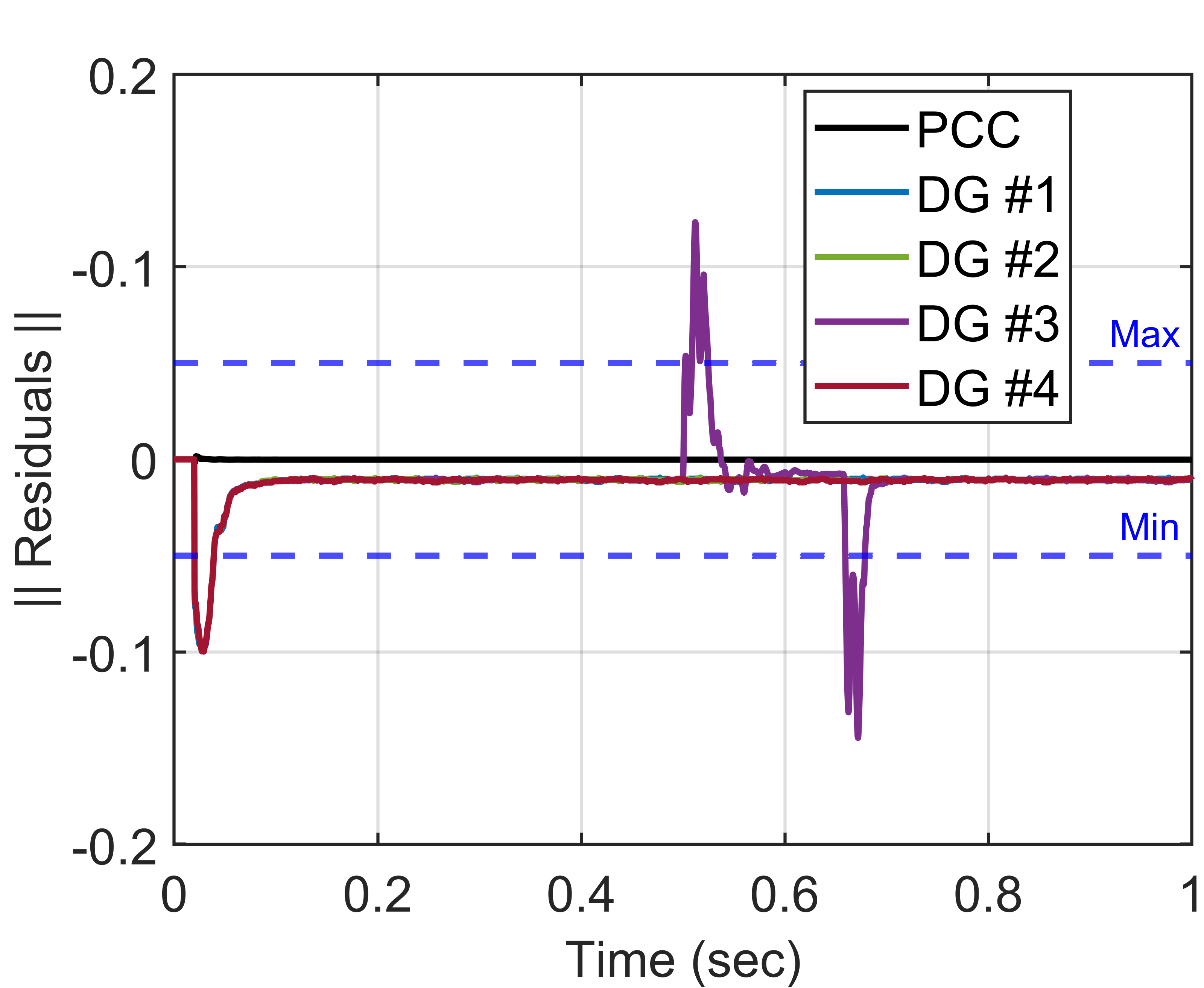}
        \label{fig:LLG}
    } 
    \subfloat[]{
        \includegraphics[width=0.23\linewidth]{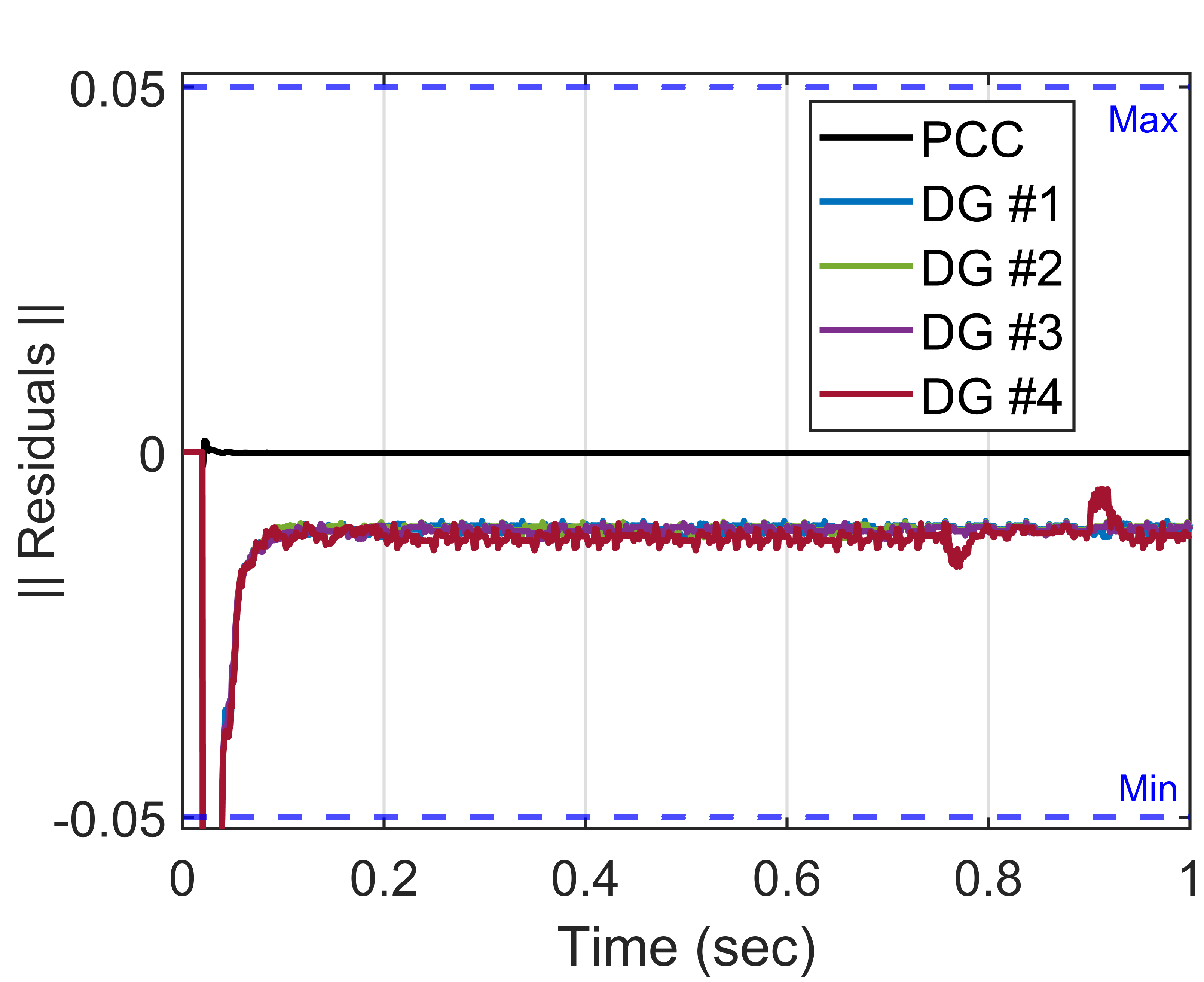}
        \label{fig:load}
    } \\
\vspace{-3mm}    
\caption[CR]{Operation of SKR triggering mechanism for different abnormal scenarios, \subref{fig:PCC}) three-phase fault at PCC, \subref{fig:control}) control input attack at DG 2 , \subref{fig:LLG}) line-to-line fault at DG 3, and \subref{fig:load}) 50\% load deviation attack at DG 4.} 
\vspace{-4mm}
\label{fig:SKR_trig}
\end{figure*}

We evaluate the efficiency of SKR alarm mechanism under diverse attack scenarios, derived from \cite{khamis2016faster}, including \emph{i)} a three-phase fault at the point-of-common-coupling (PCC) between the DG and the main grid, \emph{ii)} a control input attack, \emph{iii)} a line-to-line fault, and \emph{iv)} a 50\% load altering attack. 
Notably, the three-phase fault takes place from $0.1$ to $0.15$ $sec$, the control input attack from $0.25$ to $0.4$ $sec$, the line-to-line fault from $0.5$ to $0.65$ $sec$ and the load increase and decrease at $0.75$ and $0.95$ $sec$, respectively. The impact of the aforementioned abnormal conditions on the DG voltages and angles are presented in Fig. \ref{fig:voltAngles}. Furthermore, the SKR event detection response is demonstrated in Fig. \ref{fig:SKR_trig}. All anomalous conditions, apart from the load deviation that does not compromise the system operation (also verified by Fig. \ref{fig:voltAngles}), would trigger SKR alarms since they exceed operational thresholds.

\vspace{-3mm}
\section{Islanding Detection} \label{s:islandindDetection}
\vspace{-2mm}

This section outlines the data-driven approach for detecting anomalous grid behavior and the methodology for enforcing controlled islanding. 

\vspace{-3mm}
\vspace{-1mm}
\subsection{Preliminaries}


In ML and statistics, terms like ``independent variables'', ``predictors'', or ``features'' refer to the data used to train classifiers, while ``dependent variables,'' ``outcomes'', or ``targets'' are what they predict. These terms are interchangeable and reflect the relationship between the variables.

\vspace{-3mm}
\subsubsection{Support Vector Machine Classifiers} \label{s:SVM}


\textcolor{black}{Support vector machines (SVMs) are supervised ML algorithms used for classification and regression tasks across various scientific disciplines. To classify data, an SVM identifies hyperplanes that optimally separate the dependent variables into different classes.} 
\textcolor{black}{SVMs can be used for both linear and non-linear classification, utilizing different kernel functions to evaluate the separating hyperplane margin. Various kernel functions, such as linear, quadratic, polynomial, and radial basis, are available. This study chose a cubic and a Fine Gaussian (FGSVM) SVM kernel. Cubic SVMs have been proven to be effective classification algorithms that balance detection accuracy and speed \cite{haque2014hybrid}, while FGSVM achieve superior accuracy when dealing with multi-domain feature spaces \cite{jain2022sentiment} as in our case.}


\vspace{-3mm}
\subsubsection{Ensemble Learning Classifiers} \label{s:Ensemble}

Ensemble learning combines predictions from multiple models, or learners, to improve accuracy. Learners are classified as weak or strong based on their performance, with strong models formed by combining the outputs of weak learners. Common methods for combining predictions include boosting, stacking, and bagging \cite{zhou2021ensemble}. 
In boosting, learners are trained sequentially, with each stage focusing on reducing the errors of the previous ones. Stacking involves training different models (level-one learners) to capture diverse characteristics of the data, with a second-level model combining their outputs. Bagging trains multiple weak learners (often the same model, like decision trees) on different subsets of data, and their outputs are combined for predictions (Algorithm \ref{alg:Bagging}). 
After weak learners produce predictions decisions can be reached, typically using methods like averaging (for regression) or voting (for classification). In voting, different weights can be assigned to each learner based on accuracy. Hard voting relies on predicted classes, while soft voting uses class probabilities. In this work, we use a bagged trees, hard voting ensemble classifier due to its faster, independent (unlike boosting), parallel training process, and its speed in making predictions \cite{ardabili2019advances}.



\setlength{\textfloatsep}{0pt}
\begin{algorithm}[t]
\footnotesize
\SetAlgoNoLine
\DontPrintSemicolon
\textbf{Input:} Training Set: $\mathcal{D} = [\{\textbf{x}_{1}, y_{1}\}, \{\textbf{x}_{2}, y_{2}\}, ..., \{\textbf{x}_{k}, y_{k}\} ]$ \\

\textbf{Define} $\mathcal{N}$ = Number of training rounds    \\
\textbf{Define} $\mathcal{L}$ = Bagging Learner Model (e.g. Decision Tree)   \\

\textbf{Process:}\\
\For {$t=1, 2, ..., \mathcal{N}$}{
    $h_{t} = \mathcal{L}(\mathcal{D}, \mathcal{D}_{sub})$
}
\emph{\# Calculate Class of Target }\\
\textbf{Output:} $\mathcal{H}(\textbf{x}) = \underset{y \in \mathcal{Y}}{\argmax} \sum_{t=1}^{\mathcal{N}} \mathbb{I}(h_{t}(\textbf{x}) = y)$

\caption{Bagging Methodology.}
\label{alg:Bagging}
\end{algorithm}


\vspace{-3mm}
\subsection{Adaptive Islanding Detection Scheme}
\noindent This section outlines the methodology of the islanding scheme. 


\vspace{-3mm}
\subsubsection{Feature Vector Models}
\textcolor{black}{The islanding detection mechanism evaluates time-series data obtained from multiple sensors within the system. In this study, the measurements are sourced from the DG inverter controllers. We assume that these sensor measurements are readily accessible to the inverter controllers, and therefore, communication or transmission delays have not been accounted for in the reported results.}
For example, in the Canadian urban distribution feeder system, 54 features are used to predict anomalous behavior. These features include the DG local measurements (collected from their inverters), such as three-phase voltage, currents, and harmonic noises. Additionally, the per-unit voltages, frequencies, and angles reported at the system buses, the point-of-common-coupling (PCC), and the distribution lines are also incorporated in the feature vectors. In Eq. \eqref{eq: featureVector}, we formulate a generalized version of the input feature vector $\mathcal{\textbf{x}}$, where $\mathcal{K}$ and $\mathcal{N}$ refer to the number of measurement points and the number of DGs (providing local measurements), respectively. The voting ML classifier is then trained to fit the aggregated measurements compiled in the feature vectors.
\begin{align} \label{eq: featureVector} \vspace{-2mm}
    \begin{split}
    \mathcal{\textbf{x}} = [{}&v_{pu}^{1}, v_{pu}^{2}, ...,v_{pu}^{\mathcal{\mathcal{K}}},\delta^{1}, \delta^{2}, ..., \delta^{K}, f^{1}, f^{2}, ..., f^{\mathcal{K}}, v_{abc}^{DG1}, i_{abc}^{DG1}, thd_{abc}^{DG1}, \\ 
    {}&v_{abc}^{DG2}, i_{abc}^{DG2}, thd_{abc}^{DG2}, ..., v_{abc}^{DG\mathcal{N}}, i_{abc}^{DG\mathcal{N}}, thd_{abc}^{DG\mathcal{N}} ]    
    \end{split}
\end{align}\vspace{-2mm}
\vspace{-2mm}

These features will be mapped by ML classifiers (i.e., cubic  SVM, FGSVM, DTs, etc.) in a high-dimensional space where changes in any of these predictors will be associated with nominal or abnormal system operation. As a result, the system behavior and the evolution of its dynamic state trajectories will be directly correlated with the measured data points (within the feature vectors). 
Our methodology uses raw data allowing the dynamic update of our learning models without requiring data preconditioning steps (decompressing encoded data \cite{ryu2019convolutional, khamis2016faster}) -- adding complexity and computations -- while overcoming potential reconstruction issues for post-event analyses,  since the feature vectors might not be reversible (e.g., lossy encoding). \looseness=-1


\vspace{-3mm}
\subsubsection{Training Process}

The feature vectors are forwarded to the learning algorithms described in Sections \ref{s:SVM} and \ref{s:Ensemble} to train the islanding detection classifiers. The training of the classifiers is performed independently (and in parallel). In Table \ref{tab:learningAlg}, we compile information on the structure and the specific characteristics of the learning algorithms. 
Each corresponding algorithm enhances the fitting capabilities of the combined classifier since cubic and Gaussian kernel functions, as well as bagged decision trees, learn to predict how changes in features are reflected on the dynamic system states. 
\textcolor{black}{The learner training process is conducted offline (not real-time) and may be time-intensive depending on the feature vector dimensions and training dataset size. However, anomalous event detection is designed to be performed by the controllers in real-time.}

\begin{table}[t!]
\vspace{-5mm}
\setlength{\tabcolsep}{1.2pt}
\centering
\caption{Machine Learning Classification Algorithm Specifications}\vspace{-3mm}
\label{tab:learningAlg}
\resizebox{0.6\linewidth}{!}{
\begin{tabular}{||l|c|c||}
\hline \hline
\textbf{Learner Model} & \textbf{Parameter} & \multicolumn{1}{c||}{\textbf{Value}} \\ \hline 

 \multirow{5}{*}{Bagged Trees} & Ensemble Method & Bootstrapping Aggregation \\ \cline{2-3}
 & Learner Type & Decision Tree \\ \cline{2-3}
 & \# of Learners & $30$ \\ \cline{2-3}
 & Max \# of Splits & $200k$ \\  \cline{2-3} 
 & Miss-classification cost & unweighted \\  \hline 
 
 \multirow{3}{*}{Cubic SVM} & Kernel Function & Cubic \\ \cline{2-3}
 & Kernel Scale & $1.0$ \\ \cline{2-3}
 & Miss-classification cost & unweighted \\ \hline 
 
 \multirow{3}{*}{Fine Gaussian SVM} & Kernel Function & Fine Gaussian \\ \cline{2-3}
 & Kernel Scale & $1.8$ \\ \cline{2-3}
 & Miss-classification cost & unweighted  \\ \hline \hline
\end{tabular}}
\end{table}

\vspace{-3mm}
\subsubsection{Classification Process}

The islanding detection is a two-stage classification process. During the first stage of the classification, the three distinct learner algorithms decide independently whether the input time series data, i.e., features, correspond to nominal or abnormal operation. After the inference stage, the results of each individual classifier are forwarded to the voting algorithm, where by majority vote, the final decision for the label of the input data is made. The credibility of the final decision is validated on a rolling window basis, which comprises the second stage of the classification process. \looseness=-1

\textcolor{black}{The second stage of the classification process aggregates the results from each individual classifier and, by averaging the three latest decisions, identifies the correct class. Given that our classifiers can make real-time decisions on the input data in less than $0.022~secs$, we aggregate the inference results over three consecutive rounds forming the cumulative confidence score (Fig. \ref{fig:Detection}). If the decisions of independent classifiers are found to be consistent in the class of input measurements, i.e., decision credibility (Fig. \ref{fig:Detection}) is higher than $90\%$, then this decision is considered credible and is forwarded to the optimal islanding module, which will take the corresponding actions based on the classification results \cite{khamis2016faster}. }

If consensus between individual classifiers is not reached within the three predefined rounds, i.e., decision credibility is less than $90\%$, the rolling decision window is extended to five rounds. After the fifth round, the average decision is forwarded to the islanding module (regardless of the decision accuracy). Based on our simulation results, the proposed islanding detection mechanism can achieve an accurate decision, with a longer decision interval. However, faster and accurate detection can allow the islanding module to calculate the optimal adaptive islanding topology, minimizing power-demand imbalances and, at the same time, limiting the potential impacts of adverse events (restricting the impact propagation to just a subpart of the whole system). Algorithm \ref{alg:Classification} describes the classification and the inference of event class labels (i.e., normal or abnormal system states).

\vspace{-2mm}

\setlength{\textfloatsep}{0pt} 
\begin{algorithm}[t]
\footnotesize
\SetAlgoNoLine
\DontPrintSemicolon
\textbf{Input:} Feature Vector: $\textbf{x}_{t}$, Trained Classifiers $\mathcal{C}_{i}$ \\
\textbf{Define} $\mathcal{NR}$ = \# of decision rounds , $\mathcal{NC}$ = \# of decision classifiers  \\
\textbf{Process:}\\
$score = 0; credibility = 0;$ \\
\For {$t=1, 2, ...,  \mathcal{NR}$}{
        $score = score + \mathcal{C}_{1}(\textbf{x}_{t})$ + $\mathcal{C}_{2}(\textbf{x}_{t})$ + ... +  $\mathcal{C}_{\mathcal{NC}}(\textbf{x}_{t})$ \\
    \If{$t == 3$}{$credibility = score \div (3 \times \mathcal{NC})$ \\
    \If{$\text{credibility} \in [0, 0.1] \text{ OR } \text{credibility} \in [0.9, 1]$}{
        $\mathcal{Y}_{[k, k+t]} = round(credibility)$ \\
        break; \\
    }
    }
}
$\mathcal{Y}_{[k, k+t]} = round( score \div (\mathcal{NR} \times \mathcal{NC}))$ \emph{ \quad \# Decide Class of Event }\\

\textbf{Output:} $\mathcal{Y}_{[k, k+t]}$ 
\caption{Classification Methodology.}
\label{alg:Classification}
\end{algorithm}


\vspace{-2mm}
\section{Mixed Integer Programming Islanding Methodology} \label{s:islandingMethod} \vspace{-2mm}
This section presents the mixed integer programming (MIP) formulation for the islanding process. The methodology identifies stable islands while maximizing served load demand. The system's network topology is modeled as an undirected graph $(N, L)$, where $|N|$ nodes and $|L|$ lines represent branches connecting system buses in $B$. The edge directions are arbitrary in our method, as in other studies \cite{patsakis2019strong, zhang2014islanding}. Generators and loads are represented by $G$ and $D$, respectively. We now introduce the binary variables and constraints the optimizer must satisfy to minimize load losses.


\vspace{-4mm}
\subsection{Islanding Constraints}
\subsubsection{Generator Constraints}
The output of the generators within the system is limited by their absolute minimum and maximum ratings and their steady-state and transient response characteristics. \textcolor{black}{We are assuming that the DG inverters are operating in grid-forming mode and during an islanding event, a generator could either be completely disconnected (switched off) or operated within a feasible margin around its previous steady-state point due to its physical limitations (e.g., recovery response):} 
\vspace{-2mm}
\textcolor{black}{
\begin{equation} 
    P^{min}_{g} \leq \phi_{g} P_{g}^{t} \leq P^{max}_{g} \quad \textsf{and} \quad P_{g}^{t-1} (1 - \chi_{g})\leq \phi_{g} P_{g}^{t} \leq P_{g}^{t-1} (1 + \chi_{g}) \vspace{-2mm}
    \label{eq:genconst} 
\end{equation}
}
\textcolor{black}{
\noindent where $\phi_{g}$ is a variable that models if generator $g$ is switched on  ($\phi_{g}=1$) or off ($\phi_{g}=0$). If generator $g$ is still running at $time=t$, its power output $P_{g}^{t}$ will be within its operating, i.e., between $P^{min}_{g}$ and $P^{max}_{g}$, and ramp limits.}

\vspace{-3mm}
\subsubsection{Load Constraints}
During an islanding event the power system is sectionalized into healthy and unhealthy sections. As a result, some loads might not be able to be served due to generator capacity limitations, 
or line constraints. 
In such occasions, load shedding is performed to maintain the stability of the sectionalized areas. 
We use a load aggregation model 
to represent this relationship:
\vspace{-2mm}
\begin{equation} \vspace{-2mm}
  P_{d}^{s} = \beta_{d} P_{d}^{a} \quad s.t. \quad
  \theta_{d} \leq \beta_{d} \leq 1
    \label{eq:loadconst}
\end{equation}

\noindent where $\theta_{d}$ represents the portion of the aggregate load $d$ that is critical, i.e., has priority in being served during an islanding, and $\beta_{d}$ is the actual portion of loads being supplied. If $\beta_{d} = 1$, the entire demand is satisfied by the resources of the corresponding section of the power system to which $d$ belongs. \looseness=-1

\vspace{-4mm}
\subsubsection{Line models and constraints}

For our line modeling, we have opted for a AC power flow model. 
As a result, for our system to attain stability pre or post islanding: 
 \vspace{-2mm}
\begin{equation}\vspace{-2mm}
        \forall b \in B: \sum_{g \in G}P_{g} + \sum_{l_{i,j} \in  L^{+}}P_{l_{i,j}} - \sum_{l_{i,j} \in L^{-}}P_{l_{i,j}} = \sum_{d \in D}P_{d}^{s} \\
    \label{eq:equal}    
\end{equation}

\noindent Apart from  \eqref{eq:equal}, the power flow network should not be overloaded (especially post islanding). Thus, the real power flow line limits, i.e., $MW$ ratings, should not be exceed: 
\vspace{-2mm}
\begin{equation}\vspace{-2mm}
    P_{l_{i,j}}^{min} \leq w_{l_{i,j}} P_{l_{i,j}} \leq P_{l_{i,j}}^{max} 
    \label{eq:lineconst}
\end{equation}
\vspace{-10mm}

\subsection{Islanding Formulation}
Once a maloperating part of the system is identified, the grid topology is segregated into healthy and unhealthy network partitions. \textcolor{black}{The healthy system portion encompasses all the nodes that are operating nominally, while the unhealthy partition is defined as the maximum allowable partition proposed by our islanding strategy.} We use the binary decision variable $h_{b}$, to distinguish between buses in healthy and unhealthy partitions. Such that:
\vspace{-2mm}
\begin{equation}\vspace{-2mm}
    \forall b \in B: 
    \begin{cases}
        h_{b} = 1 \, | \, b \in B^{h}  \subseteq B \ \text{(healthy part)} \\
        h_{b} = 0 \, | \, b \in B^{u}  \subseteq B \ \text{(unhealthy part)}\\
        
    \end{cases}
    \label{eq:busPartition}    
\end{equation}

We use another binary decision variable, i.e., $w_{l_{i,j}}$, for the lines that will remain connected after partitioning the network into healthy ($B^{h}$) and unhealthy ($B^{u}$) subsets. It should be noted that disconnected lines with both ends in $B^{h}$ could also exist (e.g., if they reach their power flow limits). We denote with $L^{\bullet} \subseteq L$ the set of lines that should be disconnected as a result of physical (e.g., ratings), other limitations (e.g., connecting healthy and unhelathy islands), or could potentially cause power instabilities. The following 
describes the relation between line connectivity and binary variable $w_{l_{i,j}}$. \vspace{-2mm}
\vspace{-2mm}
\begin{equation} 
    \forall l_{i,j} \in L, i,j \in N: 
    \begin{cases}
        w_{l_{i,j}} = 1 \, | \, l_{i,j} \quad \text{(line connected)} \\
        w_{l_{i,j}} = 0 \, | \, l_{i,j} \quad \text{(line disconnected)}\\    
    \end{cases}
    \label{eq:linePartition}    
\end{equation}
        
\noindent Additional constraints exist for the lines belonging to $L^{\bullet}$. Such lines should be disconnected if they are connecting $B^{h}$ and $B^{u}$. That is, $w_{l_{i,j}} = 0 : \forall l_{i,j} \in L | h_{b_{i}} = 1, h_{b_{j}} = 0 \ \text{or} \ h_{b_{i}} = 0, h_{b_{j}} = 1$. 
Furthermore, if both ends of the line exist in $B^{h}$, the line should also be disconnected to maintain the stability of $B^{h}$. On the other hand, if both ends of the line exist in $B^{u}$, i.e., $h_{b_{i}} = 0 \ \textsf{and} \ h_{b_{j}} = 0$, the line can remain connected since the stability of $B^{u}$ cannot be guaranteed during islanding. The aforementioned constraints can be modeled by the following inequalities. \vspace{-2mm}
\begin{equation} \vspace{-2mm}
    \forall \ l_{i,j} \in L^{\bullet}:
        w_{l_{i,j}} \, \leq \, 1 - h_{b_{i}} \ \textsf{and} \ w_{l_{i,j}} \, \leq \, 1 - h_{b_{j}}
    \label{eq:lineConst1}    \vspace{-2mm}
\end{equation}
\noindent For the remaining lines, i.e., $l_{i,j} \in L \backslash L^{\bullet}$, we still have to ensure that if they connect $B^{h}$ and $B^{u}$ they need to be disconnected. However, in this case, if a line $l_{i,j}$ belongs exclusively to $B^{h}$ or $B^{u}$, it can either remain connected or disconnected without affecting the stability of the islands. These constraints are formalized by the following inequalities. \vspace{-2mm} 
\begin{equation} \vspace{-2mm}
    \forall l_{i,j} \notin  L^{\bullet}: 
        w_{l_{i,j}}  \leq 1 - h_{b_{i}} + h_{b_{j}} \; \textsf{and} \;
        w_{l_{i,j}}  \leq 1 + h_{b_{i}} - h_{b_{j}}
    \label{eq:lineConst2}    \vspace{-2mm}
\end{equation}

\subsection{Islanding Objective Function}
The objective of defensive islanding is to divide the power grid into stable islands while serving as many critical loads as possible (minimizing load imbalances). Typically, these islands are composed of generators that furnish similar angles before the islanding process is initiated, i.e., generator coherency. Apart from the optimal generator groupings efficiently meeting load demand, minimizing the number of line disconnections 
guarantees that the grid architecture (post-islanding) will be able to withstand potential future disturbances and cascading accidental or deliberate contingencies. The formulations provided in  \eqref{eq:lines}--\eqref{eq:loads} model the aforementioned  objectives. 

\begin{subequations}
    \begin{equation}
      \texttt{min} \sum_{l_{ij} \notin L^{h}} \omega_{l_{i,j}} (1 - w_{l_{i,j}})
      \label{eq:lines}
    \end{equation}  \vspace{-2mm}
    \begin{equation}
      \texttt{min} \sum_{g \in G} \omega_{g} (1 - \phi_{g})
      \label{eq:gens}
    \end{equation}  \vspace{-2mm}
    \begin{equation}
      \texttt{max} \sum_{d \in D} \beta_{d} P_{d}^{a} [h_{d} + \psi_{d}(1-h_{d})]
      \label{eq:loads}
    \end{equation}  \vspace{-2mm}
    \label{eq:objectives}
\end{subequations}

In \eqref{eq:lines} we constrain the optimizer to minimize the the number of lines that will be disconnected (through the term $1 - w_{l_{i,j}}$) and the number of generators that are going to be disconnected is similarly minimized in \eqref{eq:gens} through the $1 - \phi_{g}$ term. The weights $\omega_{l_{i,j}}$ and $\omega_{g}$ are normalized based on the maximum allowable line power flow and generator nameplate capacity to keep connected lines and generators with ``heavier'' weights. In  \eqref{eq:loads} maximizing the loads being served and minimizing the size of the unhealthy network topology are jointly optimized via the $[h_{d} + \psi_{d}(1-h_{d})]$ term. $\psi_{d}$ is a stochastic variable modeling the probability of a load $d$ belonging in the $B^{u}$ to be served. Since $0 \leq \psi_{d} \leq 1$, $\beta_{d} P_{d}^{a} \psi_{d} \leq \beta_{d} P_{d}^{a} $, thus, priority is given to the loads attached to buses in $ B^{h}$. Combining the described equations gives us the comprehensive form of the islanding objective function: \vspace{-2mm}
\begin{equation} \vspace{-2mm}
    J = \texttt{max} (\lambda_{1}\eqref{eq:loads} - \lambda_{2}\eqref{eq:lines} - \lambda_{3}\eqref{eq:gens} )
    \label{eq:objFunc} \vspace{-2mm}
\end{equation}


\noindent \textcolor{black}{
Here, $\lambda_{1}, \lambda_{2}, \lambda_{3}$ are positive weights, uniformly selected in this study, to prioritize the sub-objectives (Eq. \eqref{eq:lines}, \eqref{eq:gens}, and \eqref{eq:loads}) of the optimizer while enhancing its convergence (by limiting the size of the branch-and-bound space). The use of uniform weights ensures that the optimizer is not biased toward favoring the formation of larger islands, maintaining an equitable balance among the sub-objectives.}

\begin{figure*}[t!]
\centering
    \subfloat[]{
        \includegraphics[width=0.235\textwidth]{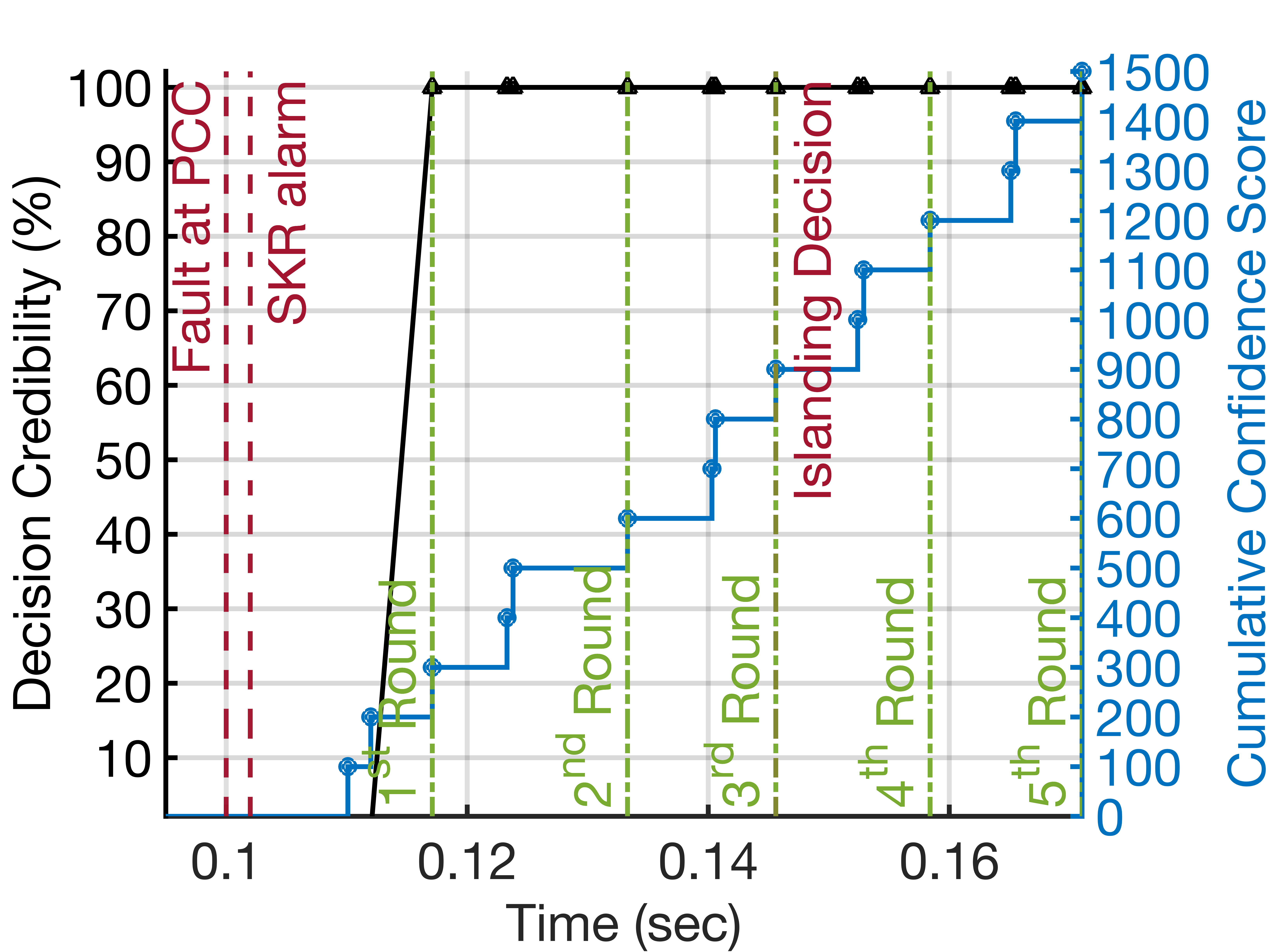}
        \label{fig:PCCdet}
    } 
    \subfloat[]{
        \includegraphics[width=0.235\linewidth]{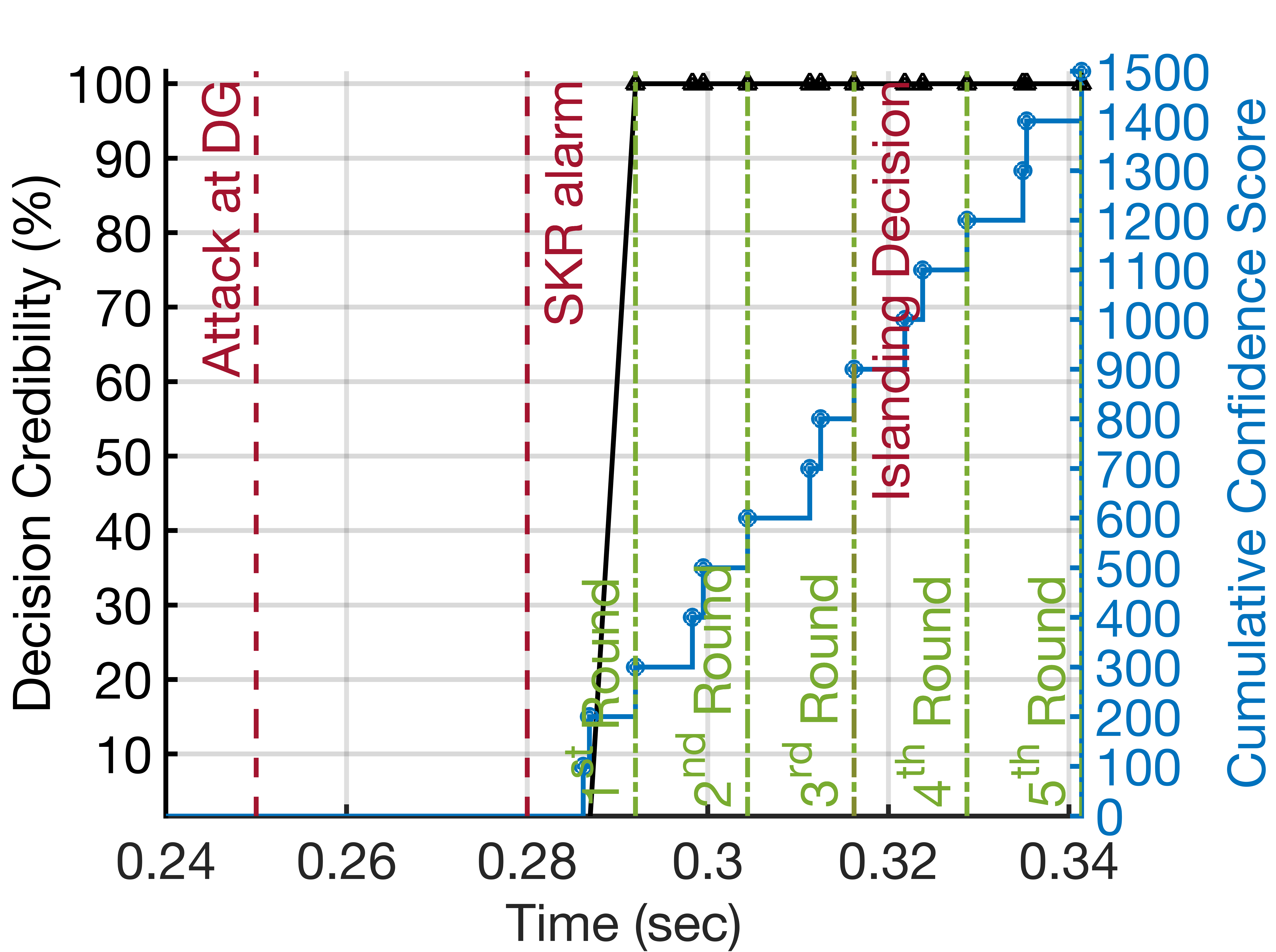}
        \label{fig:controldet}
    } 
    \subfloat[]{
        \includegraphics[width=0.235\linewidth]{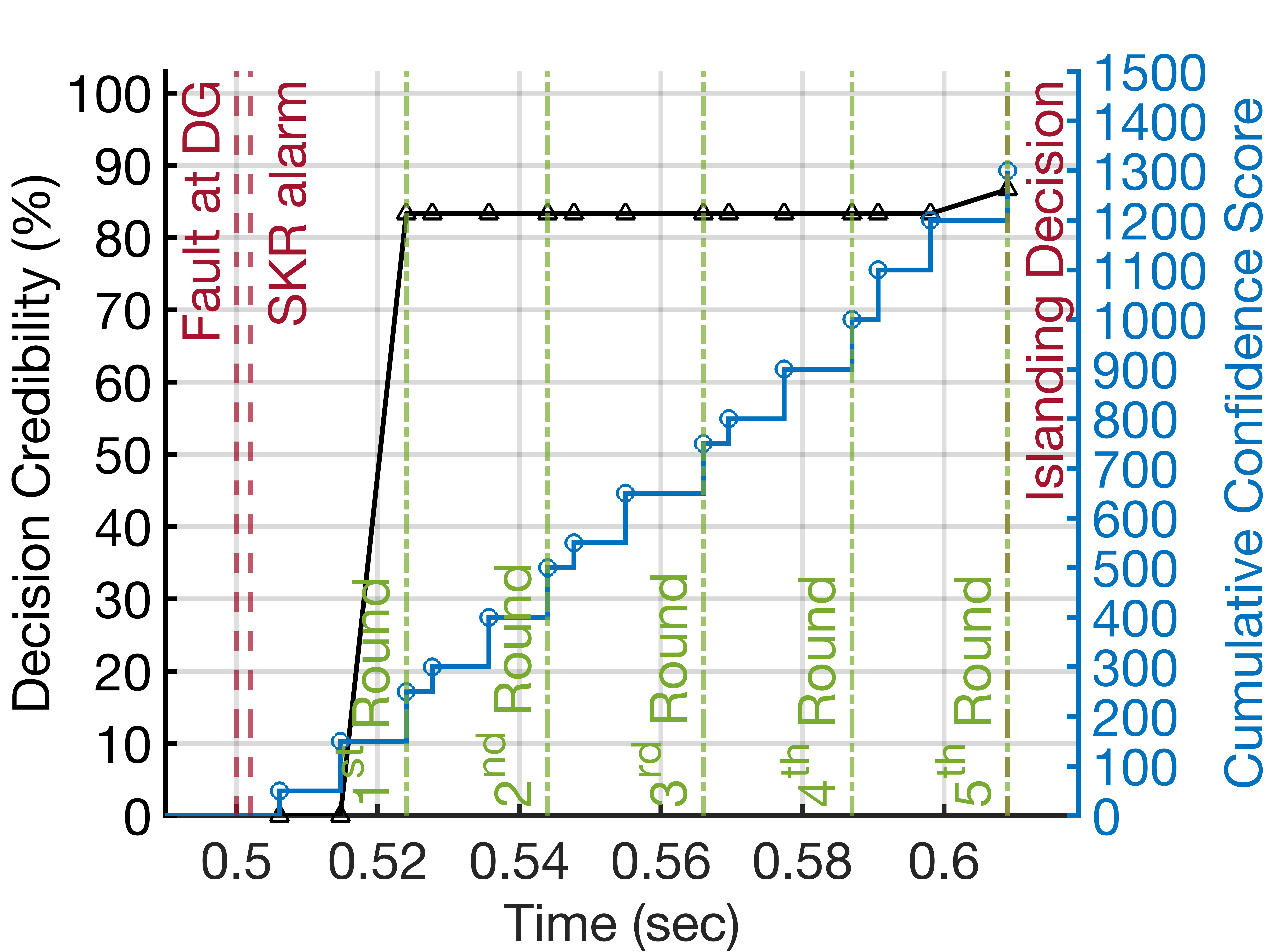}
        \label{fig:LLGdet}
    } 
    \subfloat[]{
        \includegraphics[width=0.235\linewidth]{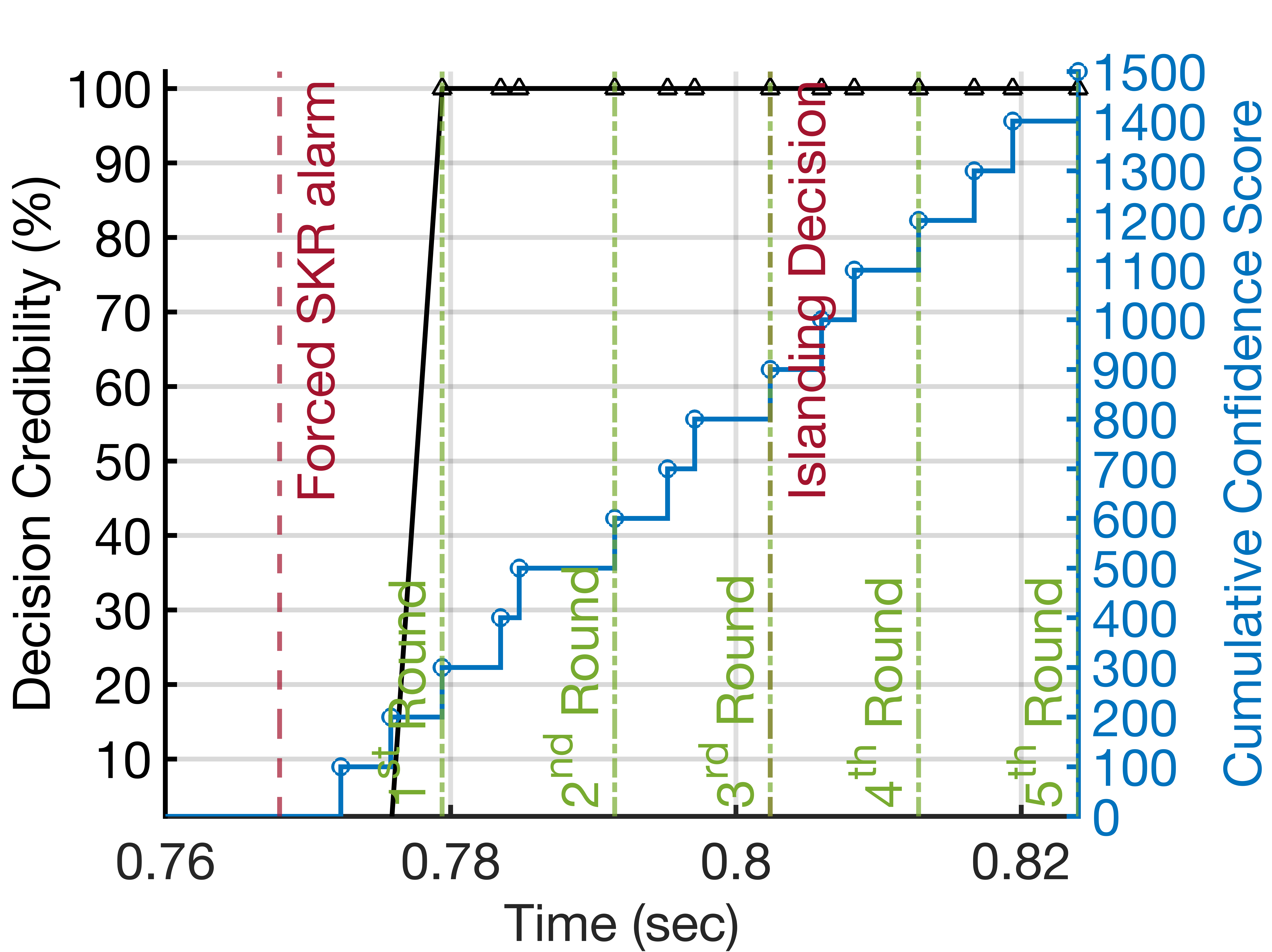}
        \label{fig:loaddet}
    } 
    
\caption[CR]{Detection mechanism response during, \subref{fig:PCCdet}) three-phase fault at PCC, \subref{fig:controldet}) control input attack at DG, \subref{fig:LLGdet}) line-to-line fault at DG, and \subref{fig:loaddet}) 50\% load deviation attack at DG.} 
\label{fig:Detection}
\end{figure*}

\vspace{-5mm}
\section{Simulation Results}\label{s:results}
\vspace{-2mm}
We demonstrate the impact of various islanding strategies on the IEEE 24- and 118-bus networks. We compare the proposed proactive islanding method, which maintains large supply-adequate islands, with traditional schemes that aggressively isolate compromised sections. The efficiency of these strategies is evaluated using MATPOWER AC power flow solutions. 

\vspace{-3mm}
\subsection{System Models} \label{s:systemModel}
This study uses the IEEE 24-bus reliability test system (RTS) and the IEEE 118-bus system to evaluate the isolation logic. The RTS consists of 24 buses, 38 branches, 17 loads, and 32 generators with a total generation capacity of 3075 MW and a net demand of 2850 MW. The 118-bus system includes 118 buses, 186 branches, 91 loads, and 54 generators, with a total generation of 4375 MW and a demand of 4242 MW. More details, including generator costs and system specifics, can be found in \cite{subcommittee1979ieee}. 

\textcolor{black}{Similar scenarios to the ones employed in \cite{khamis2016faster}, i.e., best algorithm in terms of detection speed and accuracy, 
are used to evaluate our 
methodology. Specifically, the scenarios that we aim to detect for potential islanding include: \emph{i)} fault at PCC bus, 
\emph{ii)} attack at bus (simulated control input attack on the grid-forming inverter), \emph{iii)} fault and trip of a bus, 
\emph{iv)} load altering attack (increase by $50\%$) at a bus.} 
Buses 1 and 4 for RTS-24 and 54 and 89 for the IEEE 118-bus were arbitrarily chosen, and similar detection results would occur for any other generation or load bus. The scalability of the proposed method is ensured since SKR event triggering and islanding detection are performed distributively by each system bus/agent (e.g., grid-forming DG controllers). This approach avoids single points of failure and eliminates the need for nodes with advanced computational resources. 


\vspace{-3mm}
\subsection{Islanding Detection}
In this section, we evaluate the effectiveness of the SKR event-triggered and ensemble-based detection scheme for the four scenarios introduced in Section \ref{s:SKRalarms} and Fig. \ref{fig:SKR_trig} based on \cite{khamis2016faster}. In more detail, Fig. \ref{fig:Detection} demonstrates the operation of the event-triggered detection combining the SKR alarms and voting ensemble classifier (Alg. \ref{alg:Classification}). 
\textcolor{black}{The cumulative confidence score (vertical right axis) indicates the confidence of the islanding classification results over consecutive decision rounds.
Based on the credibility percentage of the classifier outcomes drawn in each corresponding round (vertical left axis), islanding (or not) is performed if decision credibility $\in [0.9, 1.0]$, as described in Alg. \ref{alg:Classification}.} To balance decision accuracy and detection speed, islanding decisions are either made during the third or fifth round. It should be noted that in both scenarios (third or fifth round), decisions are promptly made (faster than $2~secs$), conforming with IEEE-1547. \textcolor{black}{Faster than real-time detection is essential, and our methodology guarantees that (for each round) classification can be performed in less than one cycle of a $50$/$60$ Hz system (on average), i.e., $0.02$/$0.017~secs$ respectively. Thus, allowing the final decision to be formed leveraging incoming measurements that are progressively received by the classifiers.}  

Specifically, in the scenarios illustrated in Figs. \ref{fig:PCCdet}, \ref{fig:controldet}, \ref{fig:LLGdet}, it can be seen that after the four adverse conditions are initiated (per Section \ref{s:SKRalarms}), the SKR alarms are issued. Contrary to the fault scenarios Figs. \ref{fig:PCCdet} and \ref{fig:LLGdet}, which are immediately detected, the stealthy control input attack requires more time for the alarm generation. On the other hand, in the load increase scenario,  Fig. \ref{fig:loaddet}, no SKR alarm would be generated since such an event would not compromise the system operation (Fig. \ref{fig:load}). 
\textcolor{black}{Notably, the SKR alarm would not be triggered in this scenario since there is sufficient generation to cope with the 50\% load deviation attack. To illustrate the robustness of our detection mechanism, even if an erroneous alarm signal has been generated, i.e., a false positive, we have manually issued a synthetic SKR alarm. The ensemble voting classifier is able to make a correct islanding decision even in the presence of this ``forced" alarm that we issue at $0.77 sec$. }

Regarding the islanding detection accuracy of the ensemble-based classifier, Accurate decisions are made in three decision rounds in all use cases, apart from the line-to-line fault, where two more rounds are necessary for the identification of this abnormal condition. Notably, in every scenario (even the forced one in Fig. \ref{fig:load}), accurate decisions are made. Furthermore, the per-round detection time necessary to draw decisions of our proposed methodology is superior to previously reported studies, furnished in Table \ref{tab:IslandDecisionSpeed}. \textcolor{black}{Namely, when comparing our work with the results reported by \cite{khamis2015islanding, najy2011bayesian, el2007data, gaing2004wavelet, lidula2012pattern, faqhruldin2014universal} or \cite{khamis2016faster}, which also leverages a voting classifier, we can guarantee higher detection accuracy and faster detection speeds, i.e., $0.008~secs$ average and $0.022~secs$ maximum value versus $0.021~sec$ per round (using a Windows PC with  Intel Core i5-14600 and 16 GB Memory).}

\vspace{-3mm}
\subsection{Adaptive Islanding Test Cases}
\textcolor{black}{The practicality and scalability of our approach is validated using the IEEE RTS-24 and 118 bus systems.  
As presented in Fig. \ref{fig:framework} the islanding detection and topology modules support the adaptive control islanding scheme. 
After the detection of a maloperating bus, the healthy and unhealthy partitions of the system 
should maintain the critical portion of the power demand.} \looseness=-1

\vspace{-3mm}
\subsection{IEEE RTS-24 System Test Case}
\begin{figure*}[t!]
\centering
    \subfloat[]{
        \includegraphics[width=0.27\linewidth]{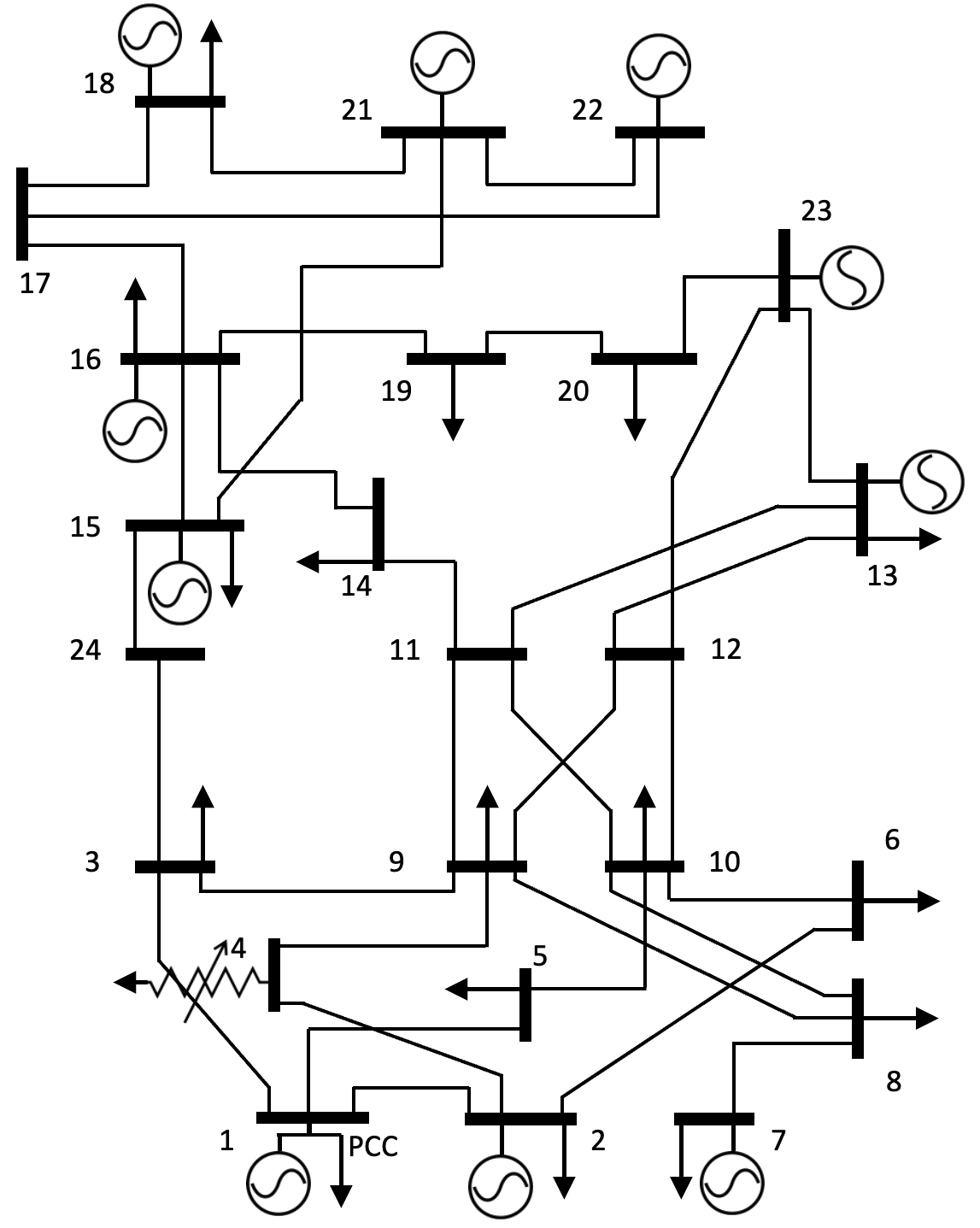}
        \label{fig:nom}
    } 
    \subfloat[]{
        \includegraphics[width=0.27\linewidth]{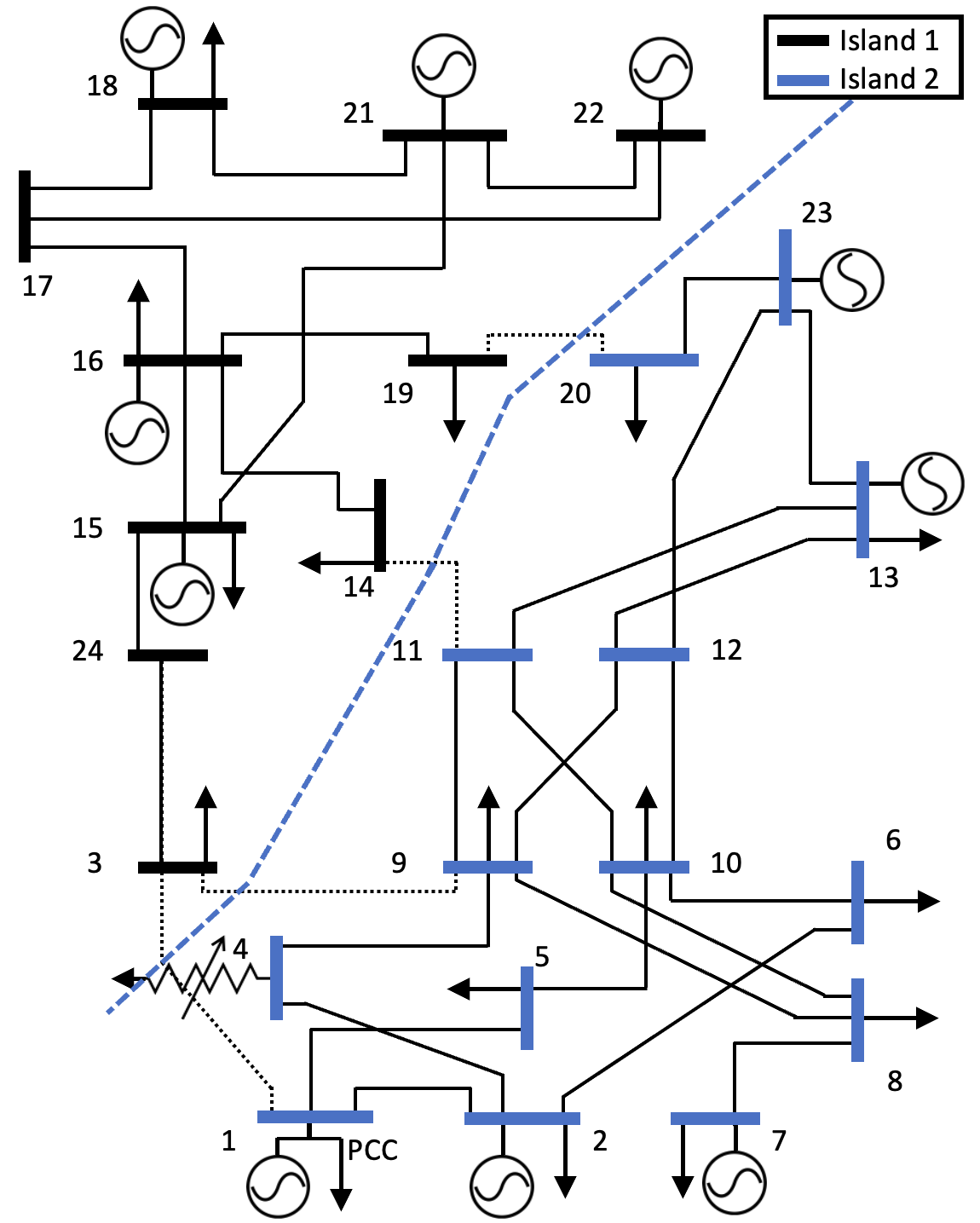}
        \label{fig:isl1}
    } 
    \subfloat[]{
        \includegraphics[width=0.27\linewidth]{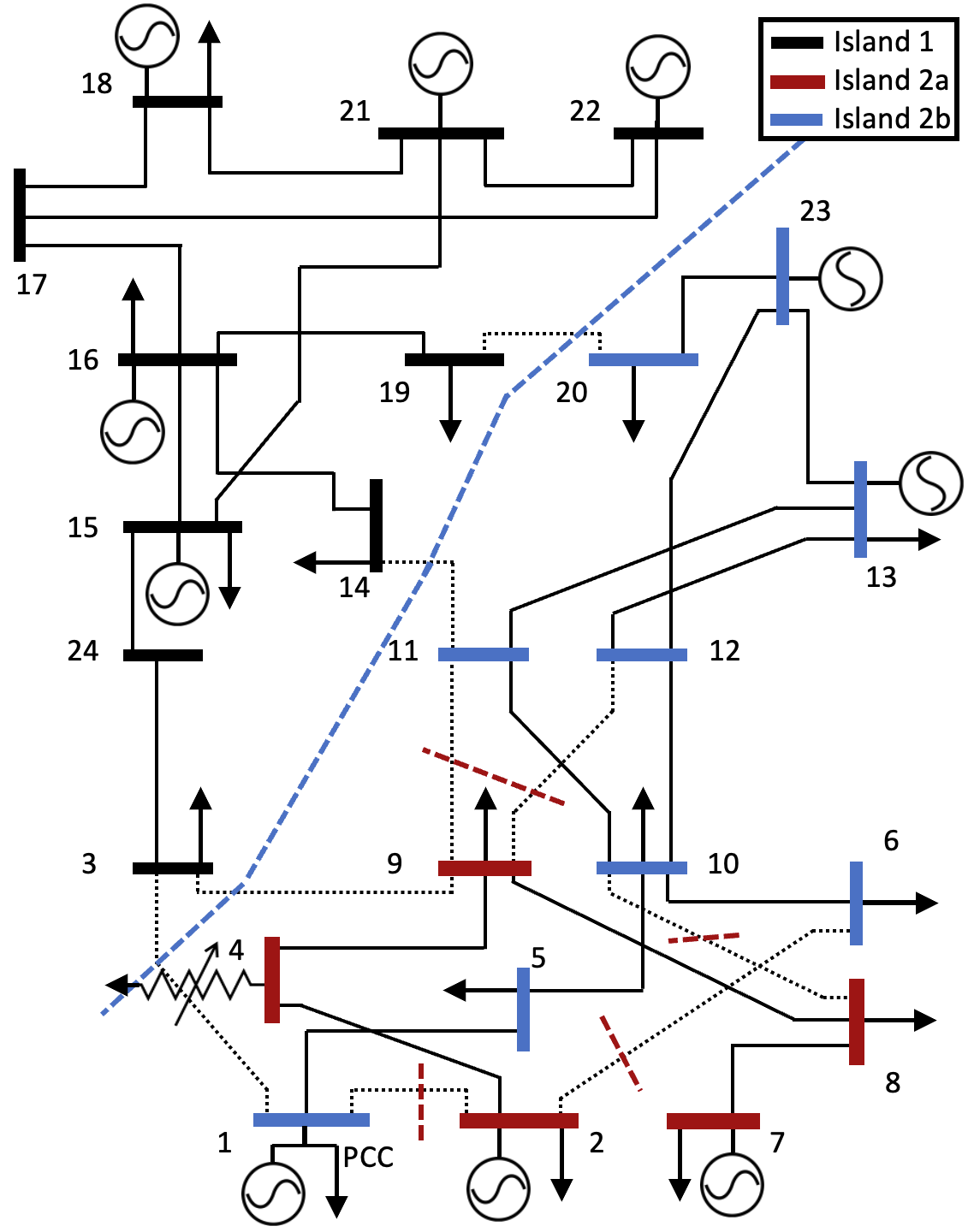}
        \label{fig:isl2}
    } \\
\vspace{-1mm}   
\caption[CR]{Modified RTS24 bus system including the controllable load at bus 4 and the PCC connection at bus 1. \subref{fig:nom}) Nominal system, i.e., pre-islanding, \subref{fig:isl1}) first step of adaptive islanding due to malicious operation at bus 4, and \subref{fig:isl2}) second step of adaptive islanding due to the persistence of the adverse conditions at bus 4, leading to load shedding. } 
\label{fig:islanding}
\end{figure*}

In the RTS-24 use case, we assume that bus 4 in Fig. \ref{fig:nom} has been compromised. After the abnormal operation of bus 4 has been detected, the adaptive controlled islanding module will select an optimal topology to sectionalize the RTS-24 system into two stable partitions. We will refer to the partition, including the anomalous agent (bus 4), as the unhealthy partition ($\forall b \in B^{u}$). The healthy partition consists of all the remaining buses, i.e., $B^{h}: \forall b \notin B^{u}$. In Fig. \ref{fig:isl1}, we present the system partitioning following the islanding methodology of Section \ref{s:islandingMethod}. We opt for larger islands instead of aggressively identifying the minimum size cut of the topology graph that will include the anomalous node (bus $4$) and is supply-adequate. 

Traditional islanding methodologies would attempt to create a section including the compromised bus $4$ as well as buses $1, 2, 7, 8, 9, 11, 12, 13$, which would satisfy the generation-demand balance of the island. Specifically, the $P_{GEN}, Q_{GEN}$ are 1245MW and 1275MVA respectively, while the demand values are $P_{DEM}, Q_{DEM}$ are 1088MW and 221MVA. However, this topology would create load imbalances in the rest of the system and increase operational costs. Other islanding topologies, including buses $1, 2, 4, 7, 8, 9$ or $1, 2, 4, 5$ would require load shedding due to load-generation imbalance but would also congest lines such as lines 1-2 and 7-8.

The proposed method identifies ``larger" stable islands while avoiding uneconomical operation, line congestion, and load shedding. The details of the nominal system, i.e., before islanding, as well as the corresponding two islands created after the detection of the malicious agent (Fig. \ref{fig:isl1}), are presented in Table \ref{tab:isl_step1}. \textcolor{black}{Leveraging MATPOWER's AC power flow solutions and the generator cost information provided in \cite{subcommittee1979ieee}, we can calculate that, before islanding, the cost of operating the whole system was \$36418.68, while after the islanding, the aggregated cost is \$36436.05 (0.05\% increase). Both islands have sufficient generation and reserves to meet current demands.} 

\begin{table}[t!]

\setlength{\tabcolsep}{1.2pt}
\centering
\caption{Step-1 of Adaptive Islanding in RTS-24 Bus System} \vspace{-1mm}
\label{tab:isl_step1}
\resizebox{0.6\linewidth}{!}{
\begin{tabular}{||l|c|c|c||}
\hline \hline
\textbf{} & \textbf{Nominal} & \textbf{Island 1} & \textbf{Island 2} \\ \hline 

{Buses} & 1-24 & {3, 14-19, 21, 21, 24} & {1, 2, 4-13, 20, 23} \\ \hline
{$P_{GEN}^{max}$} & {3075 MW} & {1470 MW} & {1605 MW} \\  \hline 
{$P_{DEM}^{max}$} & {2850 MW} & {1305 MW} & {1545 MW} \\  \hline 
{$Q_{GEN}^{max}$} & {3055 MVA} & {1470 MVA} & {1585 MVA} \\  \hline 
{$Q_{DEM}^{max}$} & {580 MVA} & {265 MVA} & {315 MVA} \\  \hline 
{Lines Disc.} & {-} & \multicolumn{2}{c||}{1-3, 3-9, 11-14, 19-20}  \\  \hline
{Cost} & {\$ 36,418.68} & {\$ 9,976.68 } & {\$ 26,459.37 } \\ \hline 
 
\hline \hline
\end{tabular}}
\end{table}

In cases where contingencies persist, the evolutionary stage of the adaptive islanding methodology will attempt to partition the unhealthy part of the system into smaller islands while meeting operational constraints. Fig. \ref{fig:isl2} demonstrates how island 2 would be further partitioned if bus 4 cannot be recovered and continues to operate abnormally. Island 2 is segregated into island 2a (unhealthy) and island 2b (healthy). The specific generation details for the aforementioned islands are summarized in Table \ref{tab:isl_step2}. It should be noted that in cases where the system constraints do not allow the design of two stable (supply-adequate) islands, the methodology prioritizes the formation of a bigger healthy partition (avoiding load shedding in this section) as defined by Eqs. \eqref{eq:lineConst1},\eqref{eq:lineConst2}, and \eqref{eq:loads}. \looseness=-1

As a result, island 2b is able to supply the load buses in the topology indicated using blue colors in Fig. \ref{fig:isl2}. However, island 2a, i.e., the persistent unhealthy partition, should perform load shedding. 
In more detail, 30.3\% of the loads located in island 2a, have to be shed due to the imbalance between $P_{GEN}$ and $P_{DEM}$, which are 502MW and 642MW, respectively. Apart from the load shedding costs, the generators in island 2a have to operate at their maximum capacity, increasing the overall cost of operating island 2 from \$26,459.37 to \$34,169.21 (29.1\% increase). \looseness=-1

\begin{table}[t!]

\setlength{\tabcolsep}{1.2pt}
\centering
\caption{Step-2 of Adaptive Islanding in RTS-24 Bus System} \vspace{-1mm}
\label{tab:isl_step2}
\resizebox{0.6\linewidth}{!}{
\begin{tabular}{||l|c|c|c||}
\hline \hline
\textbf{} & \textbf{Island 1} & \textbf{Island 2a} & \textbf{Island 2b} \\ \hline 

{Buses} & {3, 14-19, 21, 21, 24} & {2, 4, 7-9} & {1, 5, 6, 10-13, 20, 23} \\  \hline 
{$P_{GEN}^{max}$} & {1470 MW} & {502 MW} & {1103 MW}\\  \hline 
{$P_{DEM}^{max}$} & {1305 MW} & {\textbf{642} MW} & {903 MW}\\  \hline 
{$Q_{GEN}^{max}$} & {1470 MVA} & {492 MVA} & {1093 MVA}\\  \hline 
{$Q_{DEM}^{max}$} & {265 MVA} & {131 MVA} & {184 MVA}\\  \hline 
{Lines Disc.} & {1-3, 3-9, 11-14, 19-20} & \multicolumn{2}{c||}{1-2, 2-6, 8-10, 9-11, 9-12}  \\  \hline
{Load Shed} & {0\%} & {30.3\%} & {0\%}\\  \hline 
{Cost} & {\$ 9,976.68} & {\$ 8,234.64} & {\$ 25,934.57} \\ \hline 
 
\hline \hline
\end{tabular}}
\end{table}

\vspace{0mm}
\subsection{\textcolor{black}{IEEE 118 Bus System Test Case}}

\textcolor{black}{Leveraging MATPOWER's AC power flow solution and the generator cost information provided in \cite{subcommittee1979ieee}, we can calculate that, before islanding, the cost of operating the whole 118-bus system was \$140,233.19, while after the islanding, the aggregated cost is \$140,847.99 
 (0.44\% increase). Furthermore, islands 1 and 2 (comprised of islands 2a and 2b) have sufficient generation and reserves to meet the current demand. }
 
If the contingency of bus 54 is not resolved, the adaptive islanding methodology will attempt to partition the unhealthy part of the system even further. Island 2 is segregated into island 2b (unhealthy) and island 2a (healthy). 
Notably, 
the adaptive islanding does not aggressively minimize the unhealthy island 2b partition. During step-2 of the islanding procedure, the formation of the supply-adequate island 2a occurs. The generation details for the aforementioned islands are summarized in Table \ref{tab:118isl_step2}. Contrary to the RTS-24 case, sufficient generation reserves in the IEEE 118-bus system allow islands to be formed without requiring load-shedding. The islanding module will keep dividing the unhealthy partition, i.e., island 2b, into subsequent islands following the same procedure prescribed by Eq. \eqref{eq:objFunc} aiming to minimize load shedding 
while disconnecting as few line as possible, i.e., forming bigger islands.  \looseness=-1


Although load shedding was avoided in both steps of the islanding process, partitioning the system resulted in increased operational costs. Before islanding, the operational cost of the 118-bus system was \$140,233.19. After step-1, the costs of operating islands 1 and 2 are \$42,274.12 and \$98,324.54, respectively, resulting in a total cost of \$140,598.66, reflecting a 0.26\% increase. After step-2, the operational costs for islands 1, 2a, and 2b are \$42,274.12, \$7,868.03, and \$90,705.85, respectively, leading to a cumulative cost of \$140,847.99, i.e., 0.44\% increase.

\begin{table}[t!]

\setlength{\tabcolsep}{1.2pt}
\centering
\caption{Step-2 of Adaptive Islanding in 118 Bus System} \vspace{-1mm}
\label{tab:118isl_step2}
\resizebox{0.65\linewidth}{!}{
\begin{tabular}{||l|c|c|c||}
\hline \hline
\textbf{} & \textbf{Island 1} & \textbf{Island 2a} & \textbf{Island 2b} \\ \hline 

{Buses} & {82-96, 99-112} & {78-80, 97, 98} & {1-77, 81, 113-118} \\  \hline 
{$P_{GEN}^{max}$} & { 2439 MW} & { 577 MW} & { 6950.2 MW}\\  \hline 
{$P_{DEM}^{max}$} & { 946 MW} & {289 MW} & { 3007 MW}\\  \hline 
{$Q_{GEN}^{max}$} & { 4296 MVA} & { 280 MVA} & { 7201 MVA}\\  \hline 
{$Q_{DEM}^{max}$} & { 378 MVA} & { 101 MVA} & { 959 MVA}\\  \hline 
{Lines Disc.} & {77-82, 80-96, 80-99, 96-97, 98-100} & \multicolumn{2}{c||}{77-78, 77-80, 80-81}  \\  \hline
{Load Shed} & {0\%} & {0\%} & {0\%}\\  \hline 
{Cost} & {\$ 42,274.12 } & {\$ 7,868.03} & {\$ 90,705.85} \\ \hline 
 
\hline \hline
\end{tabular}}
\end{table}

\vspace{0mm}
\section{Conclusions}\label{s:conclusion}
\vspace{0mm}
\textcolor{black}{In this work, we employ islanding to answer the questions of \emph{when}, \emph{where}, and \emph{how} to deal with unexpected grid contingencies. We utilize an SKR event triggering scheme that serves as the input to the islanding detection module.} 
\textcolor{black}{In our future work, we will focus on implementing the islanding detection methodology on commercial inverter controllers and utilizing a real-time simulation testbed to evaluate its performance and quantify the communication overhead introduced by the proposed approach.}

\vspace{-5mm}
\bibliographystyle{elsarticle-num}
\bibliography{biblio}

\begin{thebibliography}{10}
\expandafter\ifx\csname url\endcsname\relax
  \def\url#1{\texttt{#1}}\fi
\expandafter\ifx\csname urlprefix\endcsname\relax\def\urlprefix{URL }\fi
\expandafter\ifx\csname href\endcsname\relax
  \def\href#1#2{#2} \def\path#1{#1}\fi

\bibitem{10870120}
I.~Zografopoulos, et~al., Cyber-physical interdependence for power system operation and control, IEEE Transactions on Smart Grid (2025) 1--1\href {https://doi.org/10.1109/TSG.2025.3538012} {\path{doi:10.1109/TSG.2025.3538012}}.

\bibitem{ospina2020trustworthy}
J.~Ospina, I.~Zografopoulos, X.~Liu, C.~Konstantinou, Trustworthy cyberphysical energy systems: Time-delay attacks in a real-time co-simulation environment, in: Proceedings of the 2020 Joint Workshop on CPS\&IoT Security and Privacy, 2020, pp. 69--69.

\bibitem{quiros2014constrained}
J.~Quir{\'o}s-Tort{\'o}s, et~al., Constrained spectral clustering-based methodology for intentional controlled islanding of large-scale power systems, IET Gen., Transm. \& Distr. 9~(1) (2014) 31--42.

\bibitem{demetriou2018real}
P.~Demetriou, M.~Asprou, E.~Kyriakides, A real-time controlled islanding and restoration scheme based on estimated states, IEEE Trans. on Power Systems 34~(1) (2018) 606--615.

\bibitem{khamis2015islanding}
A.~Khamis, H.~Shareef, A.~Mohamed, Islanding detection and load shedding scheme for radial distribution systems integrated with dispersed generations, IET GTD 9~(15) (2015) 2261--2275.

\bibitem{najy2011bayesian}
W.~K. Najy, et~al., A bayesian passive islanding detection method for inverter-based distributed generation using esprit, IEEE Trans. on Power Delivery 26~(4) (2011) 2687--2696.

\bibitem{el2007data}
K.~El-Arroudi, G.~Joos, Data mining approach to threshold settings of islanding relays in distributed generation, IEEE Trans. on power systems 22~(3) (2007) 1112--1119.

\bibitem{gaing2004wavelet}
Z.-L. Gaing, Wavelet-based neural network for power disturbance recognition and classification, IEEE Trans. on Power Delivery 19~(4) (2004) 1560--1568.

\bibitem{lidula2012pattern}
N.~W.~A. Lidula, A.~D. Rajapakse, A pattern-recognition approach for detecting power islands using transient signals—part ii: Performance evaluation, IEEE Trans. on Power Delivery 27~(3) (2012) 1071--1080.

\bibitem{faqhruldin2014universal}
O.~N. Faqhruldin, et~al., A universal islanding detection technique for distributed generation using pattern recognition, IEEE Trans. on Smart Grid 5~(4) (2014) 1985--1992.

\bibitem{khamis2016faster}
A.~Khamis, et~al., Faster detection of microgrid islanding events using an adaptive ensemble classifier, IEEE Trans. on Smart Grid 9~(3) (2016) 1889--1899.

\bibitem{liu2015principal}
X.~Liu, et~al., Principal component analysis of wide-area phasor measurements for islanding detection—a geometric view, IEEE Trans. on Power Delivery 30~(2) (2015) 976--985.

\bibitem{zografopoulos2021detection}
I.~Zografopoulos, C.~Konstantinou, Detection of malicious attacks in autonomous cyber-physical inverter-based microgrids, IEEE Trans. on Industrial Informatics (2021).

\bibitem{li2014review}
C.~Li, et~al., A review of islanding detection methods for microgrid, Renewable and Sustainable Energy Reviews 35 (2014) 211--220.

\bibitem{sun2015islanding}
Q.~Sun, et~al., An islanding detection method by using frequency positive feedback based on fll for single-phase microgrid, IEEE Trans. on Smart Grid 8~(4) (2015) 1821--1830.

\bibitem{gupta2016active}
P.~Gupta, et~al., Active rocof relay for islanding detection, IEEE Trans. on Power Delivery 32~(1) (2016) 420--429.

\bibitem{seyedi2021hybrid}
M.~Seyedi, et~al., A hybrid islanding detection method based on the rates of changes in voltage and active power for the multi-inverter systems, IEEE Trans. on Smart Grid 12~(4) (2021) 2800--2811.

\bibitem{IEEE1547}
{IEEE Std. for interconnection and interoperability of distributed energy resources with electric power systems interfaces}, IEEE Std. (2018).

\bibitem{liu2014impact}
S.~Liu, X.~Wang, P.~X. Liu, Impact of communication delays on secondary frequency control in an islanded microgrid, IEEE Trans. on Industrial Electronics 62~(4) (2014) 2021--2031.

\bibitem{8894512}
A.~{Sargolzaei}, et~al., Detection and mitigation of false data injection attacks in networked control systems, IEEE Trans. on Industrial Informatics 16~(6) (2020) 4281--4292.
\newblock \href {https://doi.org/10.1109/TII.2019.2952067} {\path{doi:10.1109/TII.2019.2952067}}.

\bibitem{zografopoulos2022time}
I.~Zografopoulos, A.~P. Kuruvila, K.~Basu, C.~Konstantinou, Time series-based detection and impact analysis of firmware attacks in microgrids, Energy Reports 8 (2022) 11221--11234.

\bibitem{haque2014hybrid}
A.~U. Haque, et~al., A hybrid intelligent model for deterministic and quantile regression approach for probabilistic wind power forecasting, IEEE Trans. on Power Systems 29~(4) (2014) 1663--1672.

\bibitem{jain2022sentiment}
A.~Jain, V.~Jain, Sentiment classification using hybrid feature selection and ensemble classifier, Journal of Intelligent \& Fuzzy Systems 42~(2) (2022) 659--668.

\bibitem{zhou2021ensemble}
Z.~H. Zhou, Ensemble learning, in: Machine learning, Springer, 2021, pp. 181--210.

\bibitem{ardabili2019advances}
S.~Ardabili, et~al., Advances in machine learning modeling reviewing hybrid and ensemble methods, in: Int'l Conf. on Global Research and Education, Springer, 2019, pp. 215--227.

\bibitem{ryu2019convolutional}
S.~Ryu, et~al., Convolutional autoencoder based feature extraction and clustering for customer load analysis, IEEE Trans. on Power Systems 35~(2) (2019) 1048--1060.

\bibitem{patsakis2019strong}
G.~Patsakis, et~al., Strong mixed-integer formulations for power system islanding and restoration, IEEE Trans. on Power Systems 34~(6) (2019) 4880--4888.

\bibitem{zhang2014islanding}
M.~Zhang, J.~Chen, Islanding and scheduling of power distribution systems with distributed generation, IEEE Trans. on Power Systems 30~(6) (2014) 3120--3129.

\bibitem{subcommittee1979ieee}
P.~M. Subcommittee, {IEEE reliability test system}, IEEE Trans. on Power Apparatus and Systems~(6) (1979) 2047--2054.

\end{thebibliography}





\end{document}